\documentclass[prd,showpacs,floatfix,amsmath,amssymb,nofootinbib,twocolumn]{revtex4-1}
\usepackage{graphicx,color,dcolumn,booktabs,bm}
\usepackage{longtable,comment,braket}
\usepackage{times}
\usepackage{overpic}
\usepackage{amssymb,tipa}
\usepackage{indentfirst}
\usepackage{feynmf}   
\usepackage{slashed}  
\usepackage{cases}
\usepackage{psfrag}
\usepackage{subfigure}
\usepackage{color}
\usepackage{multirow}

\usepackage[colorlinks, citecolor=blue,anchorcolor=red,menucolor=red,linkcolor=blue,filecolor=red,runcolor=red,urlcolor=blue,frenchlinks=red]{hyperref}

\usepackage{ulem}
\def\be{\begin{equation}}
\def\ee{\end{equation}}

\begin{document}
\title{The properties of strange quark matter under strong rotation}
\author{Fei Sun$^{1,2}$}\email{sunfei@ctgu.edu.cn}
\author{Anping Huang$^{2,3}$}\email{huanganping@ucas.ac.cn}
\affiliation{
$^1$Department of Physics, China  Three Gorges University, Yichang, 443002, China \\
$^2$Physics Department and Center for Exploration of Energy and Matter, Indiana University, 2401 N Milo B. Sampson Lane, Bloomington, Indiana 47408, USA\\
$^3$School of Nuclear Science and Technology, University of Chinese Academy of Sciences, Beijing 100049, China}

\begin{abstract}
We investigate the  rotating quark matter in the three-flavor Nambu and Jona-Lasinio (NJL) model.  The chiral condensation, spin polarization and  number susceptibility of  the light and strange quarks  are carefully studied at finite temperature without or with finite chemical potential in this model.  We find that the rotation suppresses the  chiral condensation and enhances  the first-order quark spin polarization, however for the second-order quark spin polarization and quark number susceptibility  the effect is  complicated and interesting.   When extending to the situation with finite chemical potential, we find  the angular velocity also plays a crucial role, at small angular velocity the chemical potential enhances the susceptibility, however in the  middle region of angular velocity  the effect of the chemical potential is  suppressed by the angular velocity and susceptibility  can be  changed  considerably, it can be  observed that at very low temperature in the presence of quark chemical potential the quark number susceptibility has two maxima with increasing angular velocity. Furthermore,  it is found that at  sufficiently large angular velocity the contributions  played by light quark  and strange quark to these  phenomena  are almost equal. we also explored the phase diagram in the $T$-$\omega$ plane,  we observe that there exist first order phase transitions for the rotating system and  the first order phase transition lines move toward a higher temperature for decreasing angular velocity. It is also found that the different chemical potentials change the boundary of phase diagram, and that a larger chemical potential shifts down the critical temperature. We expect these studies to be used to understand the chiral symmetry breaking  and restoration as well as probe the QCD phase transition.

\end{abstract}
\pacs{} \maketitle

\section{Introduction\label{sec1}}

QCD thermodynamics has always been the subject of intense investigations for many years, which motivates various works to try to understand it better, and we know many about the equation of state of strongly interacting matter as a function of temperature $T$ and in a limited range of quark chemical potential $\mu$.  Recently,  QCD matter under rotation is of particular interest, there  exist some interesting phenomena in rotating QCD matter, such as chiral vortical effect or chiral vortical wave \cite{D. Kharzeev, D. T. Son, D. E. Kharzeev, Y. Jiang}, which  is a key ingredient in theories that predict observable effects associated with chiral symmetry restoration and the production of false QCD vacuum states \cite{Kharzeev}. Many works can be investigated in  various rotation-related phenomena, such as noncentral heavy-ion collisions in high energy nuclear physics \cite{L. P. Csernai, F. Becattini3, Y. Jiang2, S. Shi, W. T. Deng, L. G. Pang, L. Adamczyk, F. Becattini4, X. L. Xia},  the mesonic condensation of isospin matter with rotation in hadron physics \cite{Hui Zhang}, the trapped non-relativistic bosonic cold atoms in condensed matter physics  \cite{A. L. Fetter, Urban, Iskin, R. Takahashi, J. Gooth}, the rapidly spinning neutron stars in astrophysics \cite{A. Vilenkin1,A. Vilenkin2,A. Vilenkin3,M. Kaminski,N. Yamamoto,E. Shaverin,A. L. Watts, I. A. Grenier, E. Berti}. Quark matter under rotation has  been studied
in ultra-relativistic heavy-ion off-central collisions performed at Relativistic Heavy Ion Collider (RHIC) or Large Hadron Collider (LHC) as well as lattice simulation. It is known that for the region with very low temperature and very large chemical potential there exists uncertainty in  lattice QCD due to the ``sign problem", Ref. \cite{A. Yamamoto} has calculate  the angular momenta of gluons and quarks in the rotating QCD vacuum, which would be very important for future theoretical research.

One interesting phenomenon is the quark spin polarization in noncentral collisions
of heavy ions, where  quark-gluon plasma (QGP) can be generated. And the global quark polarization could occur in the QGP  due to the large angular momentum carried by two colliding nuclei,  spin-orbit coupling can generate a spin alignment (polarization) along the direction of the system angular momentum.  This polarization provides very valuable information about the QGP properties and  can be measured experimentally with hyperons via parity-violating weak decays  \cite{Z. T. Liang, Sergei A. Voloshin, F. Becattini5, Z.T. Liang, X.G. Huang, X.G. Huang1, F. Becattini, F. Becattini1, F. Becattini2, A. Aristova, W.T. Deng, Shu Ebihara}. Experimental measurements of the $\Lambda$ hyperon polarization have been investigated  at RHIC and LHC, which would be very helpful in the study of the hottest, least viscous and most vortical-fluid ever produced both for theoretical physics and experimental physics. Recently, the global spin polarizations of $\Lambda$ and $\overline{\Lambda}$ have been measured
by the STAR collaboration in Au+Au collisions over a wide range of beam energies
$\sqrt{s_{NN}}$ = 7.7 - 200 GeV  and by ALICE collaboration in Pb+Pb collisions at
2.76 TeV and 5.02 TeV \cite{L. Adamczyk et al,J. Adam et al,S. Acharya et al}. On the other hand, theoretical research of spin polarization in the quark matter has been explored \cite{Y. Tsue1, Y. Tsue2, Y. Tsue3, H. Matsuoka1, H. Matsuoka2,X. G. Huang4,F. Becattini6,Yu Guo}, which plays an important role to explain the origin of strong magnetic field in the magnetar as well as  in changing the dynamical mass and some other phenomenon related to the chiral symmetric phase transition. It would be very interesting to take into account the influence of the rotating effect to the quark spin polarization, especially, in the case of $s$ quark matter under rotation. The study of quark spin polarization which is linked to the vorticity may help us understand the vortical nature of QGP and  the chiral dynamics of the system.

Another interesting phenomenon in non-central collisions of heavy ions is the fluctuations  and correlations of conserved charges as quantified by the corresponding susceptibilities, which
are sensitive observable quantity in relativistic heavy-ion collisions and also considered as a useful probe for QGP \cite{S. Jeon and V. Koch, V. Koch, A. Bzdak1, STAR collaboration1, S. Gupta, A. Bzdak2, A. Bazavov et al., X.F. Luo1, X. Luo, STAR collaboration2, Shuzhe Shi}.  In this paper we mainly focus on  the baryon number fluctuation which is also simply related the  quark number susceptibility (QNS). QNS serves as a signature for the QGP formation in ultra relativistic heavy-ion collision and also plays  an important factor to  probe that the QCD phase transition as well as the equation of state (EOS) of strongly interacting matter \cite{R. V. Gavai, R. V. Gavai1, S. Gupta10}.  Experimentally, various cumulants of net-kaon  multiplicity distributions of Au+Au collisions at $\sqrt{s_{NN} }$ = 7.7 - 200 GeV have been measured by the STAR experiment \cite{L. Adamczyk1, L. Adamczyk2, X. Luo2}, which are related to the thermodynamic susceptibilities. The study of QNS in lattice QCD has been interesting \cite{R. Gavai and S. Gupta, S. Borsanyi1, S. Borsanyi2, R. Bellwied, H. T. Ding}. Although susceptibilities have been studied in the past \cite{T. Kunihiro, H. Fujii and M. Ohtani, Y. Hatta and T. Ikeda, B. J. Schaefer and J. Wambach}, it has not been checked what is the influence of considering the contribution from the strange quark matter \cite{A. R. Bodmer,E. Witten,C. Alcock,C. Alcock1,J. Madsen,N. K. Glendenning1,N. K. Glendenning2,N. K. Glendenning3,H.Terazawa, H.Terazawa1,H.Terazawa2,H.Terazawa3} in the rotation system.

In this paper, encouraged by the successful description
of two-flavor QCD under rotation \cite{Y. Jiang1} in the Nambu-Jona-Lasinio  (NJL)
model, which embodies the spontaneous breaking of chiral symmetry via effective interactions between quarks, we further study the three-flavor NJL model in
the framework of quark under rotation. Since there are several flavors and colors
of quarks, several pairings are possible, which   probably lead to a great variety of  interesting phenomena. And the questions we are going to address are how the chiral condensate,  quark spin polarization and quark number susceptibility are influenced by the rotation.

Our work is organized as follows.  We first discuss the formalism of three-flavor NJL model in the presence of rotation in Section~\ref{sec2},  by using mean-field approach and the finite temperature field methods we obtain the grand potential of the fermions with rotation, and  the corresponding analytical expressions for the gap equation, spin polarization and  susceptibility of the quarks are given.  Section~\ref{sec3} presents numerical results
and discussions  in detail. Section~\ref{sec4} summarizes and concludes the paper. A brief description of fermions under  rotation is given in Appendix~\ref{A}, which
discuss  the complete set of commuting operators in cylindrical coordinates and derive the eigenstates of these operators.

\section{Formalism\label{sec2}}

We start from the NJL model, the  simplest three-flavor NJL Lagrangian for fermions without rotation is  given by \cite{M. Buballa}
\begin{eqnarray}
{\cal L} = \bar \psi \left( {i\partial _\mu  \gamma ^\mu   - m} \right)\psi  + G\sum\limits_{a = 0}^8 {\left( {\bar \psi \lambda ^a \psi } \right)^2 }+{\cal L}_{\rm det},
\end{eqnarray}
here, $\psi$ is the quark field,  $m$ is the  bare quark mass matrix, $\lambda ^a(a=1,...8) $ are the Gell-Mann matrices in flavor space with $\lambda ^0  = \sqrt {\frac{2}{3}} \boldsymbol {1}$  where $\boldsymbol {1}$ is the unit matrix in the
three-flavor space, and ${\cal L}_{\rm det}$ are given by
\begin{eqnarray}
{\cal L}_{det} = -K\{ \rm{det}[\bar \psi (1+\gamma^{5})\psi] +\rm{det}[\bar \psi (1-\gamma^{5})\psi]  \},
\end{eqnarray}
which is known as six-quark 't Hooft term, here the $\rm det$ runs over the flavor degrees of freedom and indicates the favors become connected.

\begin{widetext}

Using the mean field approximation which means that fluctuations are assumed to be small, the chiral condensates $\left\langle {\bar \psi _i \psi _j } \right\rangle$ with $i \ne j$ vanishes since it is assumed that flavor is conserved and after expanding the operators around their expectation values and neglecting the higher order fluctuations, we obtain
\begin{eqnarray}
\left( {\bar \psi _i \psi _j } \right)^2  =  - \left\langle {\bar \psi _i \psi _j } \right\rangle ^2  + 2\left\langle {\bar \psi _i \psi _j } \right\rangle \bar \psi _i \psi _j ,\left( {i = j} \right),
\end{eqnarray}
and the Lagrangian can reads
\begin{eqnarray}
\begin{array}{l}
 {\cal L} = \bar \psi\left( {i\partial _\mu  \gamma ^\mu   - M } \right)\psi- 2G\left( {\left\langle {\bar uu} \right\rangle ^2  + \left\langle {\bar dd} \right\rangle ^2  + \left\langle {\bar ss} \right\rangle ^2 } \right) +4K \left\langle {\bar uu} \right\rangle \left\langle {\bar dd} \right\rangle \left\langle {\bar ss} \right\rangle, \\
 \end{array}
\end{eqnarray}
where we have defined the dynamical quark mass $M$ as follows
\begin{eqnarray}
M_{f_{i}}  = m_{f_{i}}  - 4G\left\langle {\bar{ \psi_{f_{i}}}\psi_{f_{i}}} \right\rangle+2K\left\langle {\bar{ \psi_{f_{j}}}\psi_{f_{j}}} \right\rangle \left\langle {\bar{ \psi_{f_{k}}}\psi_{f_{k}}} \right\rangle~~~~~~~~~(i\neq j \neq k),
\end{eqnarray}
where $i,j,k$ denote the flavor of the quarks and take $u,d$ for two light flavors while $s$ for strange quark.

The condensation of the QCD matter under the presence of rotation in the 2-flavor NJL model has been investigated in  Ref. \cite{Y. Jiang1}, which exhibit interesting behavior for the paring phenomena. In the present work we will extend to study the properties of  the quark matter under rotation in 3-flavor NJL model with finite quark chemical potential. And the Lagrangian for spinor with rotation can be written  in the following way:
\begin{eqnarray}
\begin{array}{l}
 {\cal L} = \sum\bar \psi_{f}\left( {i\bar{\gamma} ^\mu (\partial _\mu+\Gamma_{\mu})- m+\gamma ^0 \mu } \right)\psi_{f} + G\sum\limits_{a = 0}^8 {\left( {\bar \psi \lambda ^a \psi } \right)^2 }+{\cal L}_{det}, \\
 \end{array}
\end{eqnarray}
here $\bar{\gamma}^\mu=e_{a}^{\ \mu} \gamma^a$ with $e_{a}^{\ \mu}$ the tetrads for spinors, $\Gamma_\mu=\frac{1}{4}\times\frac{1}{2}[\gamma^a,\gamma^b] \ \Gamma_{ab\mu}$ which is the  spinor connection, where $\Gamma_{ab\mu}=\eta_{ac}(e^c_{\ \sigma} G^\sigma_{\ \mu\nu}e_b^{\ \nu}-e_b^{\ \nu}\partial_\mu e^c_{\ \nu})$, and $G^\sigma_{\ \mu\nu}$ is the affine connection determined by $g^{\mu\nu}$. Considering the system with an angular velocity along the fixed $z$-axis, then $\vec{v}=\vec{\omega}\times \vec{x}$ and choosing $e^{a}_{\ \mu}=\delta^a_{\ \mu}+  \delta^a_{\ i}\delta^0_{\ \mu} \, v_i$ and $e_{a}^{\ \mu}=\delta_a^{\ \mu} -  \delta_a^{\ 0}\delta_i^{\ \mu} \, v_i$ (details can be found in Ref.),  then we can expand the Lagrangian to the first order of angular velocity, finally, the Lagrangian is given by

\begin{eqnarray}
{\cal L} = \bar \psi \left[ {i\gamma ^\mu  \partial _\mu   - m +\gamma ^0 \mu+ \left( {\gamma ^0 } \right)^{ - 1} \left( {\left( {\mathord{\buildrel{\lower3pt\hbox{$\scriptscriptstyle\rightharpoonup$}}
\over \omega }  \times \mathord{\buildrel{\lower3pt\hbox{$\scriptscriptstyle\rightharpoonup$}}
\over x} } \right).\left( { - i\mathord{\buildrel{\lower3pt\hbox{$\scriptscriptstyle\rightharpoonup$}}
\over \partial } } \right) + \mathord{\buildrel{\lower3pt\hbox{$\scriptscriptstyle\rightharpoonup$}}
\over \omega } .\mathord{\buildrel{\lower3pt\hbox{$\scriptscriptstyle\rightharpoonup$}}
\over S} _{4 \times 4} } \right)} \right]\psi + G\sum\limits_{a = 0}^8 {\left( {\bar \psi \lambda ^a \psi } \right)^2 }+{\cal L}_{det},
\end{eqnarray}
where  we can see as a result of rotation  the Dirac operator includes the
orbit-rotation coupling term and the spin-rotation coupling term.

The general definition of the partition function can be written as
\begin{eqnarray}
{\cal Z} = \int {D[\bar \psi ]} D[\psi ]e^{iS},
\end{eqnarray}
here, $S$ denotes the  quark action, which is the integration of the Lagrangian density $\mathcal{L}$.
After by using  the mean field approximation  and carrying out the general approach of the path integral formulation for Grassmann variables, we are now able to exactly integrate out the  fermionic fields and   obtain
\begin{eqnarray}
\begin{array}{l}
 \log {\cal Z} =  - \frac{1}{T}\int {d^3 x} \left( - 2G\left( {\left\langle {\bar uu} \right\rangle ^2  + \left\langle {\bar dd} \right\rangle ^2  + \left\langle {\bar ss} \right\rangle ^2 } \right) +4K \left\langle {\bar uu} \right\rangle \left\langle {\bar dd} \right\rangle \left\langle {\bar ss} \right\rangle \right) + \sum\limits_{f} \log \det \frac{{D_f^{ - 1} }}{T} \\
 \end{array}\label{logz},
\end{eqnarray}
here
\begin{eqnarray}
D^{ - 1}  = \gamma ^0 \left( { - i{\omega_{l}}  + \left( {n + \frac{1}{2}} \right)\omega  + \mu } \right) - M - \mathord{\buildrel{\lower3pt\hbox{$\scriptscriptstyle\rightharpoonup$}}
\over \gamma } .\mathord{\buildrel{\lower3pt\hbox{$\scriptscriptstyle\rightharpoonup$}}
\over p},
\end{eqnarray}
which is the inverse of propagator for  quarks, and
\begin{eqnarray}
\log \det \frac{{\hat D^{ - 1} }}{T} = tr\log \frac{{\hat D^{ - 1} }}{T} = \int {d^3 x} \int {\frac{{d^3 p}}{{\left( {2\pi } \right)^3 }}} \left\langle {\psi _p \left( x \right)\left| {\log \hat D^{ - 1} } \right|\psi _p \left( x \right)} \right\rangle\label{logdet}.
\end{eqnarray}
The Dirac fields can be defined in the terms of the wave functions $u(x), v(x)$
\begin{eqnarray}
\psi _p \left( x \right) = \sum\limits_{E,n,s,p} {\left( {u\left( x \right) + v\left( x \right)} \right)}\label{uv}.
\end{eqnarray}
In order to find solutions of the Dirac equation, we should firstly choose the complete set of commutating operators consists of $\hat{H}$ which can be obtained from three-flavor NJL Lagrangian, the momentum in the $z$-direction $\hat{p}_{z}$, the square of transverse momentum $\hat{p}_{t}^{2}$, the $z$-component of the total angular momentum $\hat{J}_{z}$ and the transverse helicity $\hat{h}_{t}$. The positive and negative energy solutions of the Dirac field can be determined by calculating these eigenvalue equations of the complete set of commutating operators \{$\hat{H}$, $\hat{p}_{z}$, $\hat{p}_{t}^{2}$,$\hat{J}_{z}$, $\hat{h}_{t}$\},
here, for the derivation and detailed expression of the spinor $u,v$ can refer to the  Appendix~\ref{A}, by substituting Eq. (\ref{uv}) to Eq. (\ref{logdet}) we have
\begin{eqnarray}
\log \det \frac{{\hat D^{ - 1} }}{T} = \sum\limits_{E,n,s,p} {tr\log \frac{{D_{u\left( x \right)}^{ - 1} }}{T}\int {d^3 x} } \int {\frac{{d^3 p}}{{\left( {2\pi } \right)^3 }}} \left( {\left\langle {u\left( x \right)\left| {u\left( x \right)} \right.} \right\rangle } \right)
\end{eqnarray}
\begin{eqnarray}
{\kern 1pt} {\kern 1pt} {\kern 1pt} {\kern 1pt} {\kern 1pt} {\kern 1pt} {\kern 1pt} {\kern 1pt} {\kern 1pt} {\kern 1pt} {\kern 1pt} {\kern 1pt} {\kern 1pt} {\kern 1pt} {\kern 1pt} {\kern 1pt} {\kern 1pt} {\kern 1pt} {\kern 1pt} {\kern 1pt} {\kern 1pt} {\kern 1pt} {\kern 1pt} {\kern 1pt} {\kern 1pt} {\kern 1pt} {\kern 1pt} {\kern 1pt} {\kern 1pt} {\kern 1pt} {\kern 1pt} {\kern 1pt} {\kern 1pt} {\kern 1pt} {\kern 1pt} {\kern 1pt} {\kern 1pt} {\kern 1pt} {\kern 1pt} {\kern 1pt} {\kern 1pt} {\kern 1pt} {\kern 1pt} {\kern 1pt} {\kern 1pt} {\kern 1pt} {\kern 1pt} {\kern 1pt} {\kern 1pt} {\kern 1pt} {\kern 1pt} {\kern 1pt} {\kern 1pt} {\kern 1pt} {\kern 1pt} {\kern 1pt} {\kern 1pt} {\kern 1pt} {\kern 1pt} {\kern 1pt} {\kern 1pt} {\kern 1pt} {\kern 1pt} {\kern 1pt} {\kern 1pt} {\kern 1pt} {\kern 1pt}  + \sum\limits_{E,n,s,p} {tr\log \frac{{D_{v\left( x \right)}^{ - 1} }}{T}\int {d^3 x} } \int {\frac{{d^3 p}}{{\left( {2\pi } \right)^3 }}} \left( {\left\langle {v\left( x \right)\left| {v\left( x \right)} \right.} \right\rangle } \right),
\end{eqnarray}
here, the concrete form of the ${D_u^{ - 1} }$ that has been considered the rotation and nonzero chemical potential is
\begin{eqnarray}
D_{u\left( x \right)}^{ - 1}  = \left( {\begin{array}{*{20}c}
   {\left( { - i\omega _l  + \left( {n + \frac{1}{2}} \right)\omega  + \mu } \right) - M} & { - \mathord{\buildrel{\lower3pt\hbox{$\scriptscriptstyle\rightharpoonup$}}
\over \sigma } .\mathord{\buildrel{\lower3pt\hbox{$\scriptscriptstyle\rightharpoonup$}}
\over p} }  \\
   {\mathord{\buildrel{\lower3pt\hbox{$\scriptscriptstyle\rightharpoonup$}}
\over \sigma } .\mathord{\buildrel{\lower3pt\hbox{$\scriptscriptstyle\rightharpoonup$}}
\over p} } & { - \left( { - i\omega _l  + \left( {n + \frac{1}{2}} \right)\omega  + \mu } \right) - M}  \\
\end{array}} \right),
\end{eqnarray}
which corresponding to the positive energy solution, and the concrete form for the ${D_v^{ - 1} }$ is
\begin{eqnarray}
D_{v\left( x \right)}^{ - 1}  = \left( {\begin{array}{*{20}c}
   {\left( {i\omega _l  - \left( {n + \frac{1}{2}} \right)\omega  + \mu } \right) - M} & { - \mathord{\buildrel{\lower3pt\hbox{$\scriptscriptstyle\rightharpoonup$}}
\over \sigma } .\mathord{\buildrel{\lower3pt\hbox{$\scriptscriptstyle\rightharpoonup$}}
\over p} }  \\
   {\mathord{\buildrel{\lower3pt\hbox{$\scriptscriptstyle\rightharpoonup$}}
\over \sigma } .\mathord{\buildrel{\lower3pt\hbox{$\scriptscriptstyle\rightharpoonup$}}
\over p} } & { - \left( {i\omega _l  - \left( {n + \frac{1}{2}} \right)\omega  + \mu } \right) - M}  \\
\end{array}} \right),
\end{eqnarray}
which corresponding to the negative energy solution. Here, in order to study the rotating system at finite density, we have introduced quark chemical potential $\mu$ and note that the term $\left( {n + \frac{1}{2}} \right)\omega$ in above expressions denotes the rotational polarization energy, which are very useful when we study the polarization in the following sections. By using the general methods in the finite temperature fields \cite{J. I. Kapusta}, we  obtain
\begin{eqnarray}
\begin{array}{l}
 \log \det \frac{{D_u^{ - 1} }}{T} = \beta \left( {\sqrt {M^2  + {\rm{p}}_t^2  + {\rm{p}}_z^2 } + \left( {n + \frac{1}{2}} \right)\omega } \right) \\
  \\
 {\kern 1pt} {\kern 1pt} {\kern 1pt} {\kern 1pt} {\kern 1pt} {\kern 1pt} {\kern 1pt} {\kern 1pt} {\kern 1pt} {\kern 1pt} {\kern 1pt} {\kern 1pt} {\kern 1pt} {\kern 1pt} {\kern 1pt} {\kern 1pt} {\kern 1pt} {\kern 1pt} {\kern 1pt} {\kern 1pt} {\kern 1pt} {\kern 1pt} {\kern 1pt} {\kern 1pt} {\kern 1pt} {\kern 1pt} {\kern 1pt} {\kern 1pt} {\kern 1pt} {\kern 1pt} {\kern 1pt} {\kern 1pt} {\kern 1pt} {\kern 1pt} {\kern 1pt} {\kern 1pt} {\kern 1pt} {\kern 1pt} {\kern 1pt} {\kern 1pt} {\kern 1pt} {\kern 1pt} {\kern 1pt} {\kern 1pt} {\kern 1pt} {\kern 1pt} {\kern 1pt} {\kern 1pt} {\kern 1pt} {\kern 1pt} {\kern 1pt} {\kern 1pt} {\kern 1pt} {\kern 1pt} {\kern 1pt} {\kern 1pt} {\kern 1pt} {\kern 1pt} {\kern 1pt} {\kern 1pt} {\kern 1pt} {\kern 1pt} {\kern 1pt}   + \log \left( {e^{\beta \left( {\sqrt {M^2  + {\rm{p}}_t^2  + {\rm{p}}_z^2 } + \left( {\left( {n + \frac{1}{2}} \right)\omega  - \mu } \right)} \right)}  + 1} \right) \\
  \\
 {\kern 1pt} {\kern 1pt} {\kern 1pt} {\kern 1pt} {\kern 1pt} {\kern 1pt} {\kern 1pt} {\kern 1pt} {\kern 1pt} {\kern 1pt} {\kern 1pt} {\kern 1pt} {\kern 1pt} {\kern 1pt} {\kern 1pt} {\kern 1pt} {\kern 1pt} {\kern 1pt} {\kern 1pt} {\kern 1pt} {\kern 1pt} {\kern 1pt} {\kern 1pt} {\kern 1pt} {\kern 1pt} {\kern 1pt} {\kern 1pt} {\kern 1pt} {\kern 1pt} {\kern 1pt} {\kern 1pt} {\kern 1pt} {\kern 1pt} {\kern 1pt} {\kern 1pt} {\kern 1pt} {\kern 1pt} {\kern 1pt} {\kern 1pt} {\kern 1pt} {\kern 1pt} {\kern 1pt} {\kern 1pt} {\kern 1pt} {\kern 1pt} {\kern 1pt} {\kern 1pt} {\kern 1pt} {\kern 1pt} {\kern 1pt} {\kern 1pt} {\kern 1pt} {\kern 1pt} {\kern 1pt} {\kern 1pt} {\kern 1pt} {\kern 1pt} {\kern 1pt} {\kern 1pt} {\kern 1pt} {\kern 1pt} {\kern 1pt} {\kern 1pt}  + \log \left( {e^{\beta \left( {\sqrt {M^2  + {\rm{p}}_t^2  + {\rm{p}}_z^2 } + \left( {\left( {n + \frac{1}{2}} \right)\omega  + \mu } \right)} \right)}  + 1} \right), \\
 \end{array}\label{logdetdu}
\end{eqnarray}
and
\begin{eqnarray}
\begin{array}{l}
 \log \det \frac{{D_v^{ - 1} }}{T} = \beta \left( { - \sqrt {M^2  + {\rm{p}}_t^2  + {\rm{p}}_z^2 } - \left( {n + \frac{1}{2}} \right)\omega } \right) \\
  \\
 {\kern 1pt} {\kern 1pt} {\kern 1pt} {\kern 1pt} {\kern 1pt} {\kern 1pt} {\kern 1pt} {\kern 1pt} {\kern 1pt} {\kern 1pt} {\kern 1pt} {\kern 1pt} {\kern 1pt} {\kern 1pt} {\kern 1pt} {\kern 1pt} {\kern 1pt} {\kern 1pt} {\kern 1pt} {\kern 1pt} {\kern 1pt} {\kern 1pt} {\kern 1pt} {\kern 1pt} {\kern 1pt} {\kern 1pt} {\kern 1pt} {\kern 1pt} {\kern 1pt} {\kern 1pt} {\kern 1pt} {\kern 1pt} {\kern 1pt} {\kern 1pt} {\kern 1pt} {\kern 1pt} {\kern 1pt} {\kern 1pt} {\kern 1pt} {\kern 1pt} {\kern 1pt} {\kern 1pt} {\kern 1pt} {\kern 1pt} {\kern 1pt} {\kern 1pt} {\kern 1pt} {\kern 1pt} {\kern 1pt} {\kern 1pt} {\kern 1pt} {\kern 1pt} {\kern 1pt} {\kern 1pt} {\kern 1pt} {\kern 1pt} {\kern 1pt} {\kern 1pt} {\kern 1pt} {\kern 1pt} {\kern 1pt} {\kern 1pt} {\kern 1pt}   + \log \left( {e^{\beta \left( { - \sqrt {M^2  + {\rm{p}}_t^2  + {\rm{p}}_z^2 } + \left( { - \left( {n + \frac{1}{2}} \right)\omega  - \mu } \right)} \right)}  + 1} \right) \\
  \\
 {\kern 1pt} {\kern 1pt} {\kern 1pt} {\kern 1pt} {\kern 1pt} {\kern 1pt} {\kern 1pt} {\kern 1pt} {\kern 1pt} {\kern 1pt} {\kern 1pt} {\kern 1pt} {\kern 1pt} {\kern 1pt} {\kern 1pt} {\kern 1pt} {\kern 1pt} {\kern 1pt} {\kern 1pt} {\kern 1pt} {\kern 1pt} {\kern 1pt} {\kern 1pt} {\kern 1pt} {\kern 1pt} {\kern 1pt} {\kern 1pt} {\kern 1pt} {\kern 1pt} {\kern 1pt} {\kern 1pt} {\kern 1pt} {\kern 1pt} {\kern 1pt} {\kern 1pt} {\kern 1pt} {\kern 1pt} {\kern 1pt} {\kern 1pt} {\kern 1pt} {\kern 1pt} {\kern 1pt} {\kern 1pt} {\kern 1pt} {\kern 1pt} {\kern 1pt} {\kern 1pt} {\kern 1pt} {\kern 1pt} {\kern 1pt} {\kern 1pt} {\kern 1pt} {\kern 1pt} {\kern 1pt} {\kern 1pt} {\kern 1pt} {\kern 1pt} {\kern 1pt} {\kern 1pt} {\kern 1pt} {\kern 1pt} {\kern 1pt} {\kern 1pt} {\kern 1pt}  + \log \left( {e^{\beta \left( { - \sqrt {M^2  + {\rm{p}}_t^2  + {\rm{p}}_z^2 } + \left( { - \left( {n + \frac{1}{2}} \right)\omega  + \mu } \right)} \right)}  + 1} \right), \\
 \end{array}\label{logdetdv}
\end{eqnarray}
for each flavor, here $\beta$ is the inverse temperature and the inner products of $\left\langle {u_{n,s} \left| {u_{n,s} } \right.} \right\rangle$, $\left\langle {v_{n,s} \left| {v_{n,s} } \right.} \right\rangle$  are derived with very simple expressions as follows,
\begin{eqnarray}
 \int {\frac{{d^3 p}}{{\left( {2\pi }\right)^3 }}} \left\langle {u_{n,s} \left| {u_{n,s} } \right.} \right\rangle  = \frac{1}{2}\left( {J_n \left( {p_t r} \right)^2  + J_{n + 1} \left( {p_t r} \right)^2 } \right)\label{unun},
\end{eqnarray}
\begin{eqnarray}
 \int {\frac{{d^3 p}}{{\left( {2\pi } \right)^3 }}}\left\langle {v_{n,s} \left| {v_{n,s} } \right.} \right\rangle = \frac{1}{2}\left( {J_n \left( {p_t r} \right)^2  + J_{n + 1} \left( {p_t r} \right)^2 } \right)\label{vnvn}.
\end{eqnarray}
Combining the Eqs. (\ref{logz}), (\ref{logdetdu}), (\ref{logdetdv}), (\ref{unun}) and (\ref{vnvn}) one could now  derive the expression of the grand potential for strange quark when the momentum summation turns into the integration
\begin{eqnarray}
\sum\limits_p { \to V\int {\frac{{d^3 p}}{{\left( {2\pi } \right)^3 }}} },
\end{eqnarray}
and the energy summation performs over Matsubara frequency. Then the thermodynamic grand potential   $\Omega=-\frac{T}{V} \log {\cal Z}$
has the following expression,

\begin{eqnarray}
\begin{array}{l}
\Omega {\rm{ = }}2G\left( {2{{\left\langle {\bar uu} \right\rangle }^2} + {{\left\langle {\bar ss} \right\rangle }^2}} \right) - 4K{\left\langle {\bar uu} \right\rangle ^2}\left\langle {\bar ss} \right\rangle \\
\\
 - \frac{{{N_f}{N_c}}}{{8{\pi ^2}}}\sum\limits_{n =  - \infty }^\infty  {\int_0^\Lambda  {{{\rm{p}}_t}d{{\rm{p}}_t}\int_{ - \sqrt {{\Lambda ^2} - p_t^2} }^{\sqrt {{\Lambda ^2} - p_t^2} } {d{p_z}} } \left( {\left( {{J_{n + 1}}{{({{\rm{p}}_t}r)}^2} + {J_n}{{({{\rm{p}}_t}r)}^2}} \right)} \right.}  \times \\
\\
T\left\{ {\log \left( {{e^{ - \frac{{ - {\mu _u} + \sqrt {M_u^2 + {\rm{p}}_t^2 + {\rm{p}}_z^2}  - \left( {n + \frac{1}{2}} \right)\omega }}{T}}} + 1} \right) + \log \left( {{e^{\frac{{ - {\mu _u} + \sqrt {M_u^2 + {\rm{p}}_t^2 + {\rm{p}}_z^2}  - \left( {n + \frac{1}{2}} \right)\omega }}{T}}} + 1} \right)} \right.\\
\\
\left. { + \log \left( {{e^{ - \frac{{{\mu _u} + \sqrt {M_u^2 + {\rm{p}}_t^2 + {\rm{p}}_z^2}  - \left( {n + \frac{1}{2}} \right)\omega }}{T}}} + 1} \right) + \log \left( {{e^{\frac{{{\mu _u} + \sqrt {M_u^2 + {\rm{p}}_t^2 + {\rm{p}}_z^2}  - \left( {n + \frac{1}{2}} \right)\omega }}{T}}} + 1} \right)} \right\}\\
\\
 - \frac{{{N_c}}}{{8{\pi ^2}}}\sum\limits_{n =  - \infty }^\infty  {\int_0^\Lambda  {{{\rm{p}}_t}d{{\rm{p}}_t}\int_{ - \sqrt {{\Lambda ^2} - p_t^2} }^{\sqrt {{\Lambda ^2} - p_t^2} } {d{p_z}} } \left( {\left( {{J_{n + 1}}{{({{\rm{p}}_t}r)}^2} + {J_n}{{({{\rm{p}}_t}r)}^2}} \right)} \right.}  \times \\
\\
T\left\{ {\log \left( {{e^{ - \frac{{ - {\mu _s} + \sqrt {M_s^2 + {\rm{p}}_t^2 + {\rm{p}}_z^2}  - \left( {n + \frac{1}{2}} \right)\omega }}{T}}} + 1} \right) + \log \left( {{e^{\frac{{ - {\mu _s} + \sqrt {M_s^2 + {\rm{p}}_t^2 + {\rm{p}}_z^2}  - \left( {n + \frac{1}{2}} \right)\omega }}{T}}} + 1} \right)} \right.\\
\\
\left. { + \log \left( {{e^{ - \frac{{{\mu _s} + \sqrt {M_s^2 + {\rm{p}}_t^2 + {\rm{p}}_z^2}  - \left( {n + \frac{1}{2}} \right)\omega }}{T}}} + 1} \right) + \log \left( {{e^{\frac{{{\mu _s} + \sqrt {M_s^2 + {\rm{p}}_t^2 + {\rm{p}}_z^2}  - \left( {n + \frac{1}{2}} \right)\omega }}{T}}} + 1} \right)} \right\}.
\end{array}
\end{eqnarray}
\end{widetext}

Here, the isospin symmetry has been considered for the light quarks, so $m_{d}=m_{u}, \mu_{d}=\mu_{u}$, $N_{f}=2, N_{c}=3$ and $\Lambda$ is the three-momentum. And $\mu_{u}$ is the
chemical potential for the $u$ or $d$ quark as well as $\mu_{s}$ is that for $s$ quark. When the isospin symmetry is satisfied, the dynamical quark masses are simplified to give:
\begin{eqnarray}
{M_u} = {m_u} + \left( {2K\left\langle {\bar ss} \right\rangle  - 4G} \right)\left\langle {\bar uu} \right\rangle,
\end{eqnarray}
\begin{eqnarray}
{M_s} = {m_s} - 4G\left\langle {\bar ss} \right\rangle  + 2K{\left\langle {\bar uu} \right\rangle ^2}\;.
\end{eqnarray}

We have discussed the evaluating of grand potential of quark under rotation in detail in the previous section. In this section, we list our final analytical expressions of the gap equation, quark spin polarization and quark number susceptibility in the situation  with rotation.
Firstly, we consider the gap equation which will be required to  minimize the grand potential, the values  are determined by solving the stationary condition, namely, $\frac{{\partial \Omega }}{{\partial \left\langle {\bar uu} \right\rangle }} = \frac{{\partial \Omega }}{{\partial \left\langle {\bar ss} \right\rangle }} = 0$, and the detailed expressions for the stationary condition are listed  as follows,
\begin{widetext}
\begin{eqnarray}
\begin{array}{l}
8G\left\langle {\bar uu} \right\rangle  - 8K\left\langle {\bar uu} \right\rangle \left\langle {\bar ss} \right\rangle  - \frac{{{N_c}}}{{8{\pi ^2}}}\sum\limits_{n =  - \infty }^\infty  {\int_0^\Lambda  {{{\rm{p}}_t}d{{\rm{p}}_t}\int_{ - \sqrt {{\Lambda ^2} - p_t^2} }^{\sqrt {{\Lambda ^2} - p_t^2} } {d{p_z}} } \left( {\left( {{J_{n + 1}}{{({{\rm{p}}_t}r)}^2} + {J_n}{{({{\rm{p}}_t}r)}^2}} \right)} \right.} \times \\ \\
\left\{ {\frac{{4{N_f}{e^{\frac{{{\mu _u}}}{T}}}\left( {{e^{\frac{{2n\omega  + \omega }}{T}}} - {e^{\frac{{2\sqrt {M_u^2 + p_t^2 + p_z^2} }}{T}}}} \right)\left( {2G - K\left\langle {\bar ss} \right\rangle } \right){M_u}}}{{\sqrt {M_u^2 + p_t^2 + p_z^2} \left( {{e^{\frac{{\sqrt {M_u^2 + p_t^2 + p_z^2} }}{T}}} + {e^{\frac{{2n\omega  + 2{\mu _u} + \omega }}{{2T}}}}} \right)\left( {{e^{\frac{{\sqrt {M_u^2 + p_t^2 + p_z^2}  + {\mu _u}}}{T}}} + {e^{\frac{{\left( {n + \frac{1}{2}} \right)\omega }}{T}}}} \right)}}} \right.\\ \\
\left. { - \frac{{8K{e^{\frac{{{\mu _s}}}{T}}}\left( {{e^{\frac{{2n\omega  + \omega }}{T}}} - {e^{\frac{{2\sqrt {M_s^2 + p_t^2 + p_z^2} }}{T}}}} \right)\left\langle {\bar uu} \right\rangle {M_s}}}{{\sqrt {M_s^2 + p_t^2 + p_z^2} \left( {{e^{\frac{{\sqrt {M_s^2 + p_t^2 + p_z^2} }}{T}}} + {e^{\frac{{2n\omega  + 2{\mu _s} + \omega }}{{2T}}}}} \right)\left( {{e^{\frac{{\sqrt {M_s^2 + p_t^2 + p_z^2}  + {\mu _s}}}{T}}} + {e^{\frac{{\left( {n + \frac{1}{2}} \right)\omega }}{T}}}} \right)}}} \right\} = 0,
\end{array}
\end{eqnarray}\\

\begin{eqnarray}
\begin{array}{l}
4G\left\langle {\bar ss} \right\rangle  - 4K{\left\langle {\bar uu} \right\rangle ^2} - \frac{{{N_c}}}{{8{\pi ^2}}}\sum\limits_{n =  - \infty }^\infty  {\int_0^\Lambda  {{{\rm{p}}_t}d{{\rm{p}}_t}\int_{ - \sqrt {{\Lambda ^2} - p_t^2} }^{\sqrt {{\Lambda ^2} - p_t^2} } {d{p_z}} } \left( {\left( {{J_{n + 1}}{{({{\rm{p}}_t}r)}^2} + {J_n}{{({{\rm{p}}_t}r)}^2}} \right)} \right.} \times \\ \\
\left\{ { - \frac{{4K{N_f}{e^{\frac{{{\mu _u}}}{T}}}\left( {{e^{\frac{{2n\omega  + \omega }}{T}}} - {e^{\frac{{2\sqrt {M_u^2 + p_t^2 + p_z^2} }}{T}}}} \right)\left\langle {\bar uu} \right\rangle {M_u}}}{{\sqrt {M_u^2 + p_t^2 + p_z^2} \left( {{e^{\frac{{\sqrt {M_u^2 + p_t^2 + p_z^2} }}{T}}} + {e^{\frac{{2n\omega  + 2{\mu _u} + \omega }}{{2T}}}}} \right)\left( {{e^{\frac{{\sqrt {M_u^2 + p_t^2 + p_z^2}  + {\mu _u}}}{T}}} + {e^{\frac{{\left( {n + \frac{1}{2}} \right)\omega }}{T}}}} \right)}}} \right.\\ \\
\left. {\frac{{8G{e^{\frac{{{\mu _s}}}{T}}}\left( {{e^{\frac{{2n\omega  + \omega }}{T}}} - {e^{\frac{{2\sqrt {M_s^2 + p_t^2 + p_z^2} }}{T}}}} \right){M_s}}}{{\sqrt {M_s^2 + p_t^2 + p_z^2} \left( {{e^{\frac{{\sqrt {M_s^2 + p_t^2 + p_z^2} }}{T}}} + {e^{\frac{{2n\omega  + 2{\mu _s} + \omega }}{{2T}}}}} \right)\left( {{e^{\frac{{\sqrt {M_s^2 + p_t^2 + p_z^2}  + {\mu _s}}}{T}}} + {e^{\frac{{\left( {n + \frac{1}{2}} \right)\omega }}{T}}}} \right)}}} \right\} = 0.
\end{array}
\end{eqnarray}
\end{widetext}

Here we are going to study the quark spin polarization which can be defined as  taking the partial derivative of minus grand potential with respect to angular velocity, and we introduce the following quark spin polarization as in Ref. \cite{F. Becattinia}
\begin{eqnarray}
\mathcal{P}_{1}=\frac{\partial(\frac{-\Omega}{T^{4}})}{\partial(\frac{\omega}{T})},
\end{eqnarray}
\begin{eqnarray}
\mathcal{P}_{2}=\frac{\partial^{2}(\frac{-\Omega}{T^{4}})}{\partial(\frac{\omega}{T})^{2}},
\end{eqnarray}
such definition ensures dimensionless polarization, then  we list the detailed expression of the first-order polarization and second-order polarization  as follows,

\begin{widetext}

\begin{eqnarray}
\begin{array}{l}
{{\cal P}_1} = \frac{{{N_c}}}{{8{\pi ^2}{T^3}}}\sum\limits_{n =  - \infty }^\infty  {\int_0^\Lambda  {{{\rm{p}}_t}d{{\rm{p}}_t}\int_{ - \sqrt {{\Lambda ^2} - p_t^2} }^{\sqrt {{\Lambda ^2} - p_t^2} } {d{p_z}} } \left( {\left( {{J_{n + 1}}{{({{\rm{p}}_t}r)}^2} + {J_n}{{({{\rm{p}}_t}r)}^2}} \right)} \right.} (2n + 1) \times \\ \\
\left\{ {\frac{{{N_f}\sinh \left( {\frac{{ - 2\sqrt {M_u^2 + {\rm{p}}_t^2 + {\rm{p}}_z^2}  + 2n\omega  + \omega }}{{2T}}} \right)}}{{\left( {\cosh \left( {\frac{{ - 2\sqrt {M_u^2 + {\rm{p}}_t^2 + {\rm{p}}_z^2}  + 2n\omega  + \omega }}{{2T}}} \right) + \cosh \left( {\frac{{{\mu _u}}}{T}} \right)} \right)}} + \frac{{\sinh \left( {\frac{{ - 2\sqrt {M_s^2 + {\rm{p}}_t^2 + {\rm{p}}_z^2}  + 2n\omega  + \omega }}{{2T}}} \right)}}{{\left( {\cosh \left( {\frac{{ - 2\sqrt {M_s^2 + {\rm{p}}_t^2 + {\rm{p}}_z^2}  + 2n\omega  + \omega }}{{2T}}} \right) + \cosh \left( {\frac{{{\mu _s}}}{T}} \right)} \right)}}} \right\},
\end{array}
\end{eqnarray}
\begin{eqnarray}
\begin{array}{l}
{{\cal P}_2} = \frac{{{N_c}}}{{64{\pi ^2}{T^3}}}\sum\limits_{n =  - \infty }^\infty  {\int_0^\Lambda  {{{\rm{p}}_t}d{{\rm{p}}_t}\int_{ - \sqrt {{\Lambda ^2} - p_t^2} }^{\sqrt {{\Lambda ^2} - p_t^2} } {d{p_z}} } \left( {\left( {{J_{n + 1}}{{({{\rm{p}}_t}r)}^2} + {J_n}{{({{\rm{p}}_t}r)}^2}} \right)} \right.} {(2n + 1)^2} \times \\ \\
\left\{ {{N_f}\left( {{\rm{sec}}{{\rm{h}}^2}\left( {\frac{{2{\mu _u} - 2\sqrt {M_u^2 + {\rm{p}}{{\rm{t}}^2} + {\rm{p}}{{\rm{z}}^2}}  + 2n\omega  + \omega }}{{4T}}} \right) + {\rm{sec}}{{\rm{h}}^2}\left( {\frac{{ - 2{\mu _u} - 2\sqrt {M_u^2 + {\rm{p}}{{\rm{t}}^2} + {\rm{p}}{{\rm{z}}^2}}  + 2n\omega  + \omega }}{{4T}}} \right)} \right)} \right.\\ \\
\left. { + \left( {{\rm{sec}}{{\rm{h}}^2}\left( {\frac{{2{\mu _s} - 2\sqrt {M_s^2 + {\rm{p}}{{\rm{t}}^2} + {\rm{p}}{{\rm{z}}^2}}  + 2n\omega  + \omega }}{{4T}}} \right) + {\rm{sec}}{{\rm{h}}^2}\left( {\frac{{ - 2{\mu _s} - 2\sqrt {M_s^2 + {\rm{p}}{{\rm{t}}^2} + {\rm{p}}{{\rm{z}}^2}}  + 2n\omega  + \omega }}{{4T}}} \right)} \right)} \right\}.
\end{array}
\end{eqnarray}\\
\end{widetext}

Next, we will investigate how much the rotation affects the baryon number fluctuations, these fluctuations can be quantified by the susceptibilities and the QNS is  defined through the Taylor expansion coefficients of the pressure over the chemical potential \cite{R.V. Gavai and S. Gupta1,R.V. Gavai and S. Gupta2, R.V. Gavai and S. Gupta3,C.R. Allton et al.,M. Cheng et al.,S. Bors anyi et al.,HotQCD collaboration}:
 \begin{eqnarray}
 \chi_{n}=\frac{\partial^{n}(\frac{P}{T^{4}})}{\partial(\frac{\mu}{T})^{n}},
 \end{eqnarray}
here we focus on  the second order derivative of pressure with respect to $\mu$ , due to  symmetry all the odd susceptibilities vanishes when $\mu\rightarrow 0$ (note, in the context of lattice calculations the susceptibilities are often defined as dimensionless quantities), using the relation  of pressure $P=- \Omega$, the actual calculation of the susceptibilities is straightforward and the detailed result is
\begin{widetext}
\begin{eqnarray}
\begin{array}{l}
\chi _2^f = \frac{{{N_c}}}{{16{\pi ^2}{T^3}}}\sum\limits_{n =  - \infty }^\infty  {\int_0^\Lambda  {{{\rm{p}}_t}d{{\rm{p}}_t}\int_{ - \sqrt {{\Lambda ^2} - p_t^2} }^{\sqrt {{\Lambda ^2} - p_t^2} } {d{p_z}} } \left( {\left( {{J_{n + 1}}{{({{\rm{p}}_t}r)}^2} + {J_n}{{({{\rm{p}}_t}r)}^2}} \right)} \right.}  \times \\ \\
\left( {{\rm{sec}}{{\rm{h}}^2}\left( {\frac{{2{\mu _f} - 2\sqrt {M_f^2 + {\rm{p}}{{\rm{t}}^2} + {\rm{p}}{{\rm{z}}^2}}  + 2n\omega  + \omega }}{{4T}}} \right) + {\rm{sec}}{{\rm{h}}^2}\left( {\frac{{ - 2{\mu _f} - 2\sqrt {M_f^2 + {\rm{p}}{{\rm{t}}^2} + {\rm{p}}{{\rm{z}}^2}}  + 2n\omega  + \omega }}{{4T}}} \right)} \right),
\end{array}
\end{eqnarray}
\end{widetext}
here $f=u, d, s$.
With the analytical expressions given above, we show our numerical results in the next section.

\section{Numerical results and discussions \label{sec3}}

In this section we will present our numerical results for  the gap equation, quark spin polarization and quark number susceptibility.  In our previous analytic expressions, the $z$-angular-momentum quantum number $n=0,\pm1,\pm2...$  in principle we should sum all the values of $n$,  fortunately, these expressions converge so fast that it is enough for us to sum $n$ from $-5$ to $5$, it should be noted that in order that the causality of a rigidly rotating system is maintained we should make sure that the local velocity is smaller than the light velocity, namely, the condition $\omega r<1$ should be considered in all the calculations, and for simplicity, we  take the same value of  $r$ as in Ref. \cite{Y. Jiang1}. Since any uniformly rotating system
should be spatially bounded, it has been expected that the presence of boundaries can modify the properties of the rotating system  \cite{H. L. Chen,S. Ebihara,M. N. Chernodub1,M. N. Chernodub2,M. N. Chernodub3}, indeed,  this is only true when  the angular velocity $\omega$ is much smaller than the inverse of the system's size \cite{Xinyang Wang} as well as  our  discussion is mainly devoted to the bulk properties of rotating system,  so in our analytic derivation we   ignore the finite volume
boundary effect  and  we leave it as   our further study. In our calculations, for simplicity, we have let the chemical potential $\mu_{u/d}=\mu_{s}=\mu$, and the  input parameters in the NJL are the coupling constants $G$, $K$, the quark masses $m_{u/d}$, $m_{s}$ and the three-momentum cutoff $\Lambda$.  Then, in this context we use the second case in Table 1. of Ref. \cite{H.Kohyama}, that $m_u = m_d = 0.005~{\rm{GeV}},~ m_s  = 0.1283~{\rm{GeV}},
 ~G = 3.672~{\kern 1pt} {\rm{GeV}}^{ - 2},~K = 59.628~{\kern 1pt} {\rm{GeV}}^{ - 5},
 ~\Lambda  = 0.6816~{\rm{GeV}}$, which have been estimated by the fitting in light of the following observations: $m_{\pi}=138$ MeV,~$f_{\pi}=92$ MeV,~$m_{K}=495$ MeV and $m_{\eta'}=958$ MeV.

\begin{figure}[!htbp]
\begin{center}
\setlength{\unitlength}{1mm}
\centering
\includegraphics[height=5cm]{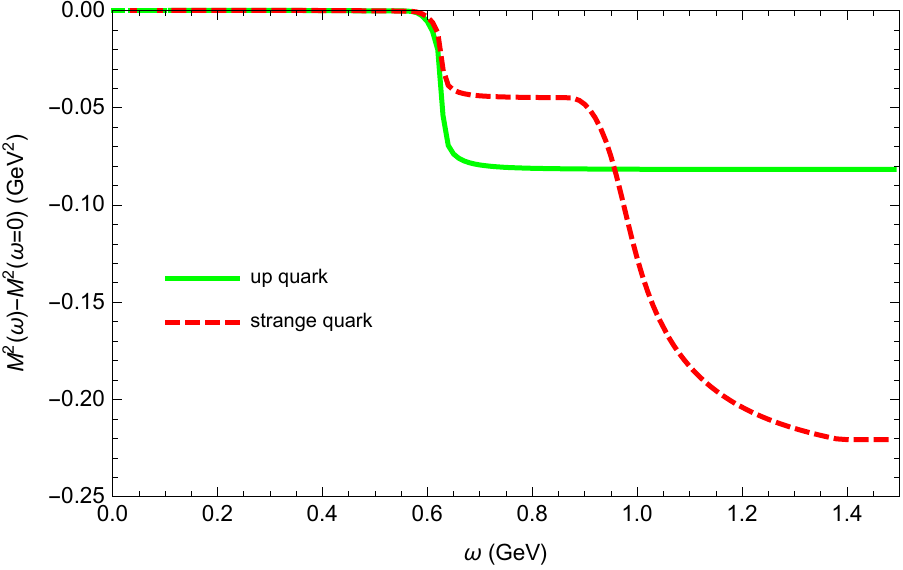}
\caption[]{(Color online) Differences of squared gap masses between the case at $\omega\neq 0$ and $\omega=0$ with both $\mu=0$ and $T=0$ for up quark and strange quark as a function of $\omega$.}
\label{plot-gapequationsquare.pdf}
\end{center}
\end{figure}

Let us first discuss the results at chemical potential equals zero.
We investigate the chiral condensation in rotating matter under the three-flavor NJL model, especially we consider the contribution from $s$ quark.
It is equivalent to study the gap equations since the gap equations correspond to the coupling of the quarks naked masses with the associated chiral condensates, firstly, in the  plot of Fig. \ref{plot-gapequationsquare.pdf}, we show the differences of squared gap masses between the case  at nonzero angular velocity to those at zero angular velocity, $M^{2}(\omega)-M^{2}(\omega=0)$ with zero temperature and  zero chemical potential for the up quark and strange quark, respectively, we found that  the squared gap mass differences of all quarks decrease with increasing angular velocity, and at large angular velocity  there is a sudden drop down for the squared gap mass differences, it obvious that  the squared gap mass differences of lighter quark  is more affected by the angular velocity, which decreases faster while that of strange quark decreases slower with angular velocity, this can be interpreted as that the quark with large current mass is less affected by the rotation.    Then  we plot the gap equation  as a function of  $\omega$  with  different values of $T$  in Fig. \ref{gapmu0.pdf} respectively for $u$ and $s$ quark. As one can see, both $u$ quark and $s$ quark gap equations show that the rotation has a suppression effect for the chiral condensation. It is clearly seen that at all temperatures  the gap equations  of  both $u$ and $s$ decrease with increasing $\omega$ and at very low temperature the chiral condensate experiences a first-order transition when $\omega$ exceeds a certain value. It is very interesting when comparing the different flavor situation, in the left panel of Fig. \ref{gapmu0.pdf} shows that ${M_u}$ experiences a first-order transition around $\omega=0.6$  GeV at  $T=0.01$ GeV,  in the right panel of Fig. \ref{gapmu0.pdf}  we observe that ${M_s}$  also experiences a sharp change  around $\omega=0.6$ GeV but changes not so much their mass compared to the light quark situation, due to the coupling between the different flavors. In addition, we also found that  for the strange quark there is a  fast drop from a high value for the quark effective mass to a small value around $\omega=1.0$  GeV. From the figure we can see  the role of the   $\omega$  as well as $T$  are very important parameters for crossover or first-order transition. For the high temperatures the chiral condensate vanishes with the increasing $\omega$ via a smooth crossover, and the temperature effect becomes weak with increasing the value of $\omega$.  As $\omega$  further increases the gap equations for the both  decrease more slowly, and both approach their naked masses.  This can be interpreted as a hint that the chiral  phase transitions for the $s, u$ do not occur at the same angular velocity with the same temperature. It is found that the chiral condensation of $u$ quark has produced results almost in agreement with those suggested in literature \cite{Y. Jiang}, but with a slight difference due to they adopted the parameters in Ref.  \cite{S. P. Klevansky}.

\begin{widetext}

\begin{figure}[!htbp]
\begin{center}
\setlength{\unitlength}{1mm}
\centering
\includegraphics[height=8cm]{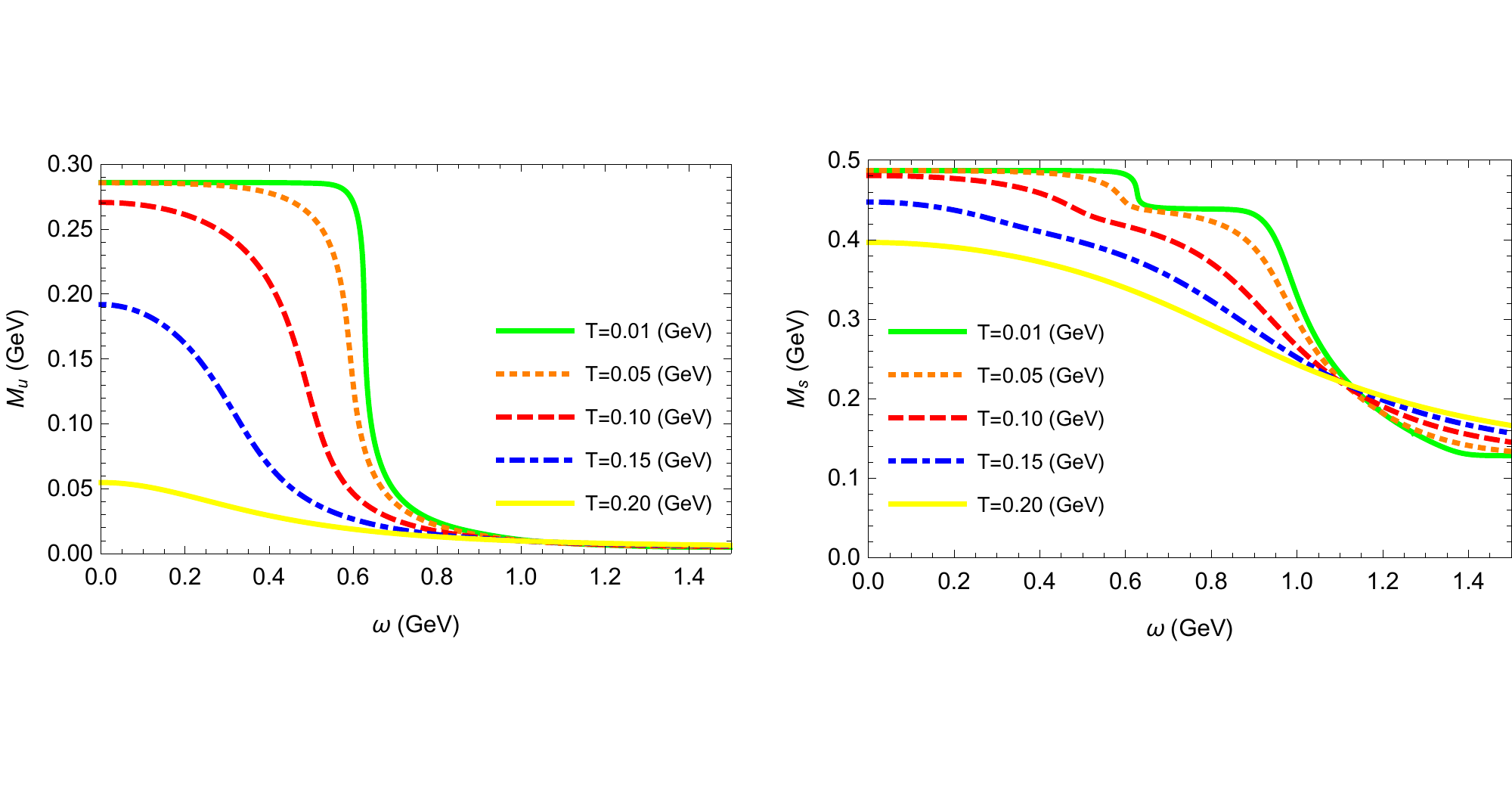}
\caption[]{(Color online) The mean field mass gap $M_{u}$ and $M_{s}$   as a function of $\omega$ for several values of $T$ with $\mu=0$.}
\label{gapmu0.pdf}
\end{center}
\end{figure}

\begin{figure}[!htbp]
\begin{center}
\setlength{\unitlength}{1mm}
\centering
\includegraphics[height=8cm]{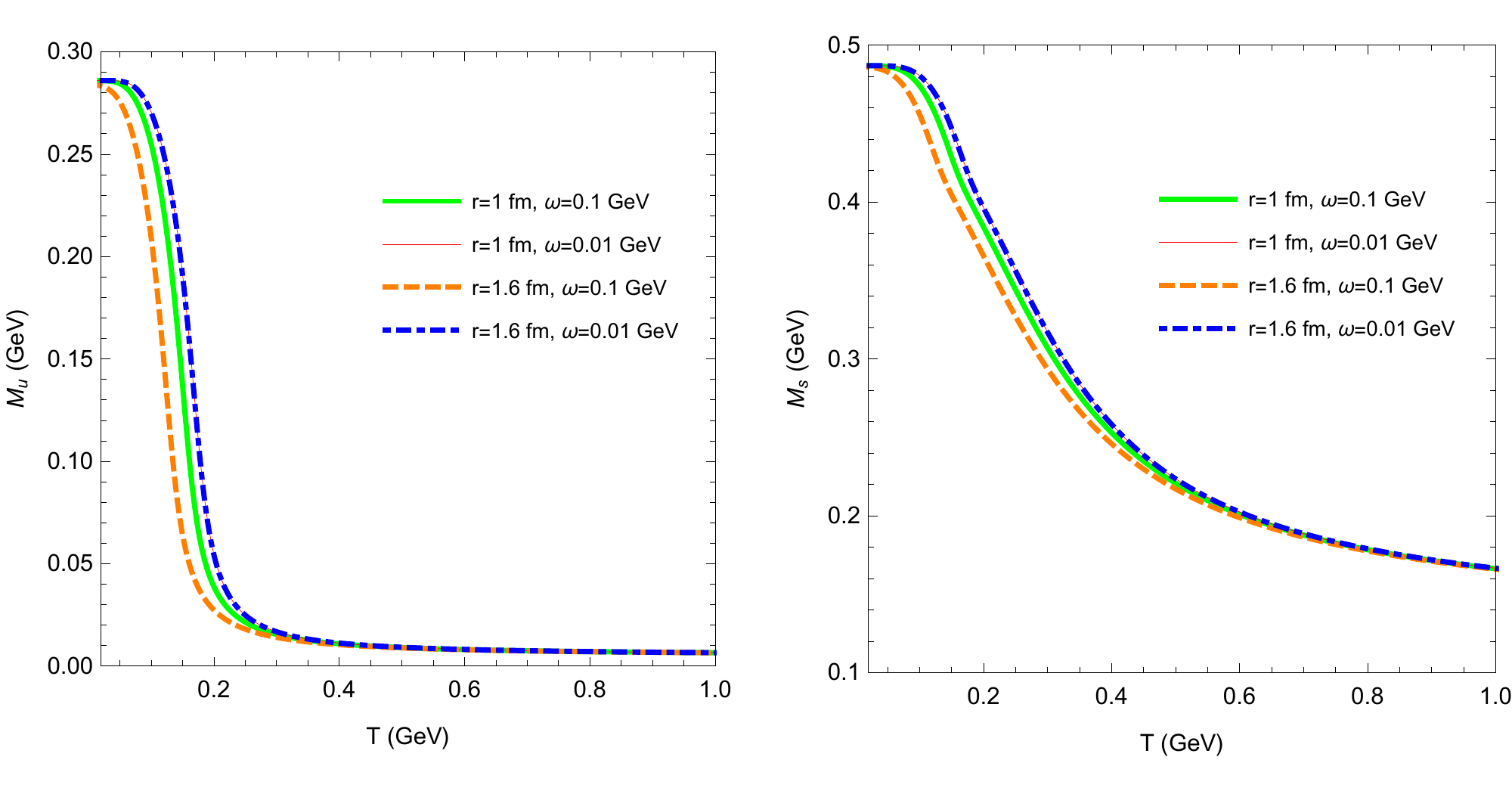}
\caption[]{(Color online) The mean field mass gap $M_{u}$ and $M_{s}$   as a function of $T$ for several values of $r$ and $\omega$ with $\mu=0$.}
\label{usMt.pdf}
\end{center}
\end{figure}
\end{widetext}

We now turn to a more realistic physical environment, in Fig. \ref{usMt.pdf} we plot the gap equation as a function of $T$ with fixed angular velocity $\omega$ and radius $r$   for $u$ and $s$ quark, respectively. In order to enable a more realistic comparison to
experimental data in the future, here  without losing generality, we assume that $\omega=0.01, 0.1$ GeV and $r=1, 1.6$ fm. From the figure we can  see clearly that there exist some interesting behaviors in the range of $ T=0.1\sim 0.4$ GeV. The left hand side of the Fig. \ref{usMt.pdf} shows that the $u$ quark gap mass decreases with increasing temperature, the mass gap $M_{u}$ falls sharply in the range of $ T=0.15\sim0.2$ GeV, however, it is a smooth behavior which means that at low $T$, small $\omega$ and large radius the quark condensate experiences a fast crossover  transition. Therefore, in this region ($ T=0.15\sim0.2$ GeV), there occurs not a true phase transition with corresponding critical temperature, but rather a pseudo-phase transition (cross-over). The right hand side of the Fig. \ref{usMt.pdf} shows that for the chosen parameters the $s$ quark gap mass decreases with increasing temperature via a crossover transition and the chiral condensate gradually vanishes with increasing temperature.
From the  Fig. \ref{usMt.pdf} one can also see clearly that, for a fixed value of $r$ the effect of rotation suppresses the effective quark mass and for a fixed value of  small $\omega$ we can observe that the effective quark mass becomes smaller at larger radius. Here we just want to   capture the essential physics of QCD matter under rotation, indeed, in realistic physical environment requires a more detailed investigation.

\begin{figure}[!htbp]
\begin{center}
\setlength{\unitlength}{1mm}
\centering
\includegraphics[height=5cm]{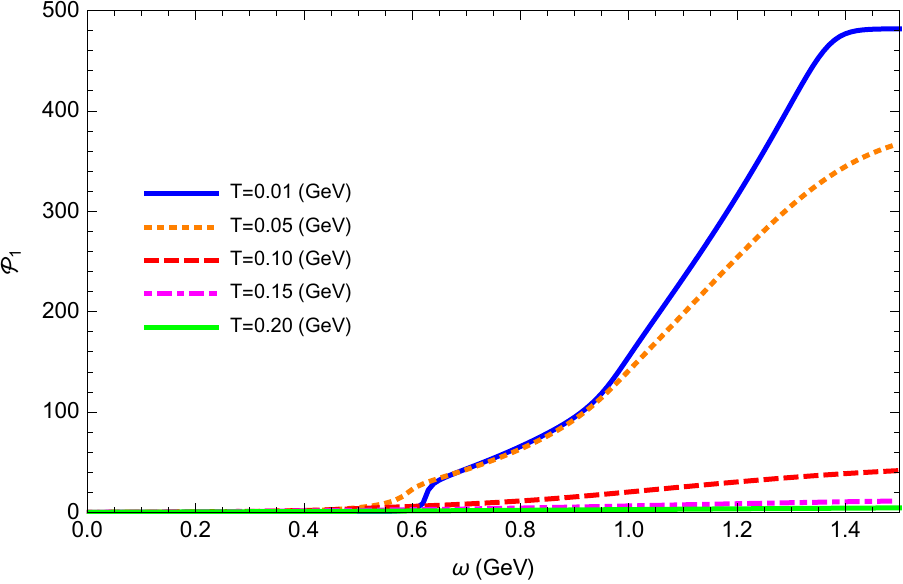}
\caption[]{(Color online) Study of the first-order spin polarizations of the rotating system according to $\omega$ for several temperature $T$  with $\mu=0$, here the result of $T =0.01$ GeV is divided by 100.}
\label{p1polarization.pdf}
\end{center}
\end{figure}

The first-order  spin polarizations of the rotating system as a function of $\omega$ for several temperatures with chemical potential equals zero are shown in Fig. \ref{p1polarization.pdf}, to be clear, the result of $T =0.01$ GeV is divided by 100. It is very clear to see that the angular velocity has  a strong influence on   the quark first-order  spin polarization as well as the temperature is also  important to the polarization of the quark. From the figure it is observed  that   the rotation system may induce a large polarization. At all temperatures the quark spin polarization increases with increasing angular velocity  for all quarks. At low temperature the quark first-order  spin polarization increases very rapidly in a certain angular velocity window and then increases very slowly, while at high temperature the quark first-order  spin polarization increases almost linearly. And at very low temperature  an  interesting phenomenon of  the jump of the quark first-order spin polarization  can be observed around $\omega\sim 0.6$ GeV for the rotating system in the figure.  This jump of the first-order spin polarization is a hint for the first order phase transition to occur, and this distinguishing feature of the spin polarization  may provide valuable insights for predicting the first order phase transition in experiments.

\begin{figure}[!htbp]
\setlength{\unitlength}{1mm}
\centering
\includegraphics[height=5cm]{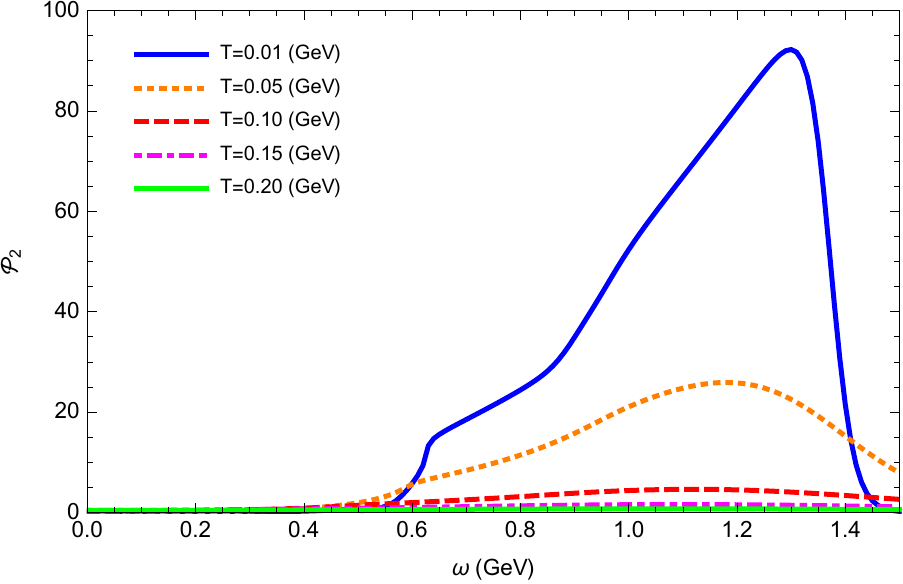}
\caption[]{(Color online) Study of the second-order spin polarizations  of the rotating system  according to $\omega$ for several  temperature $T$  with $\mu=0$.}
\label{p2polarization.pdf}
\end{figure}

\begin{widetext}

\begin{figure}[!htbp]
\setlength{\unitlength}{1mm}
\centering
\includegraphics[height=8cm]{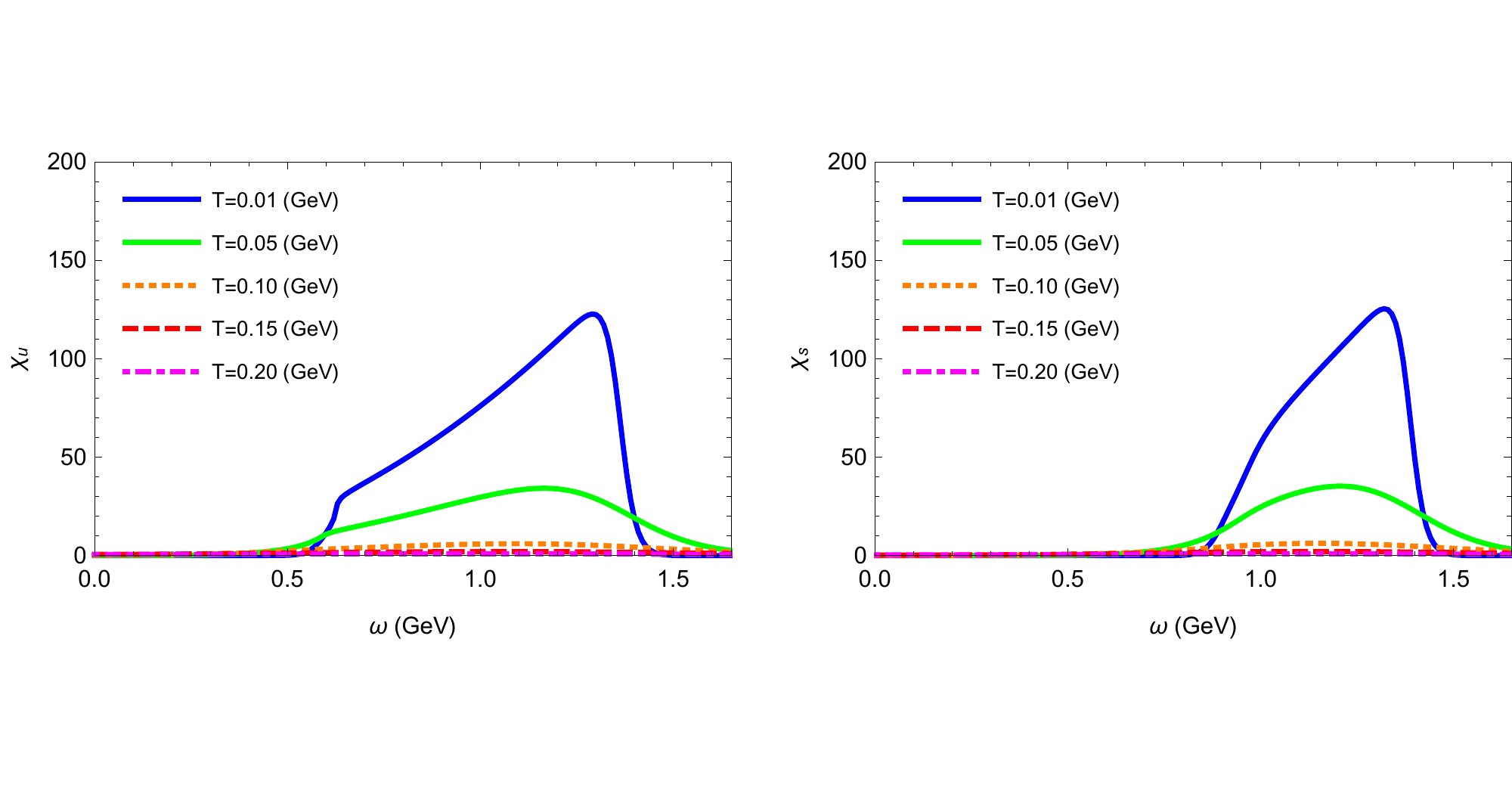}
\caption[]{(Color online) Susceptibilities of $u$ and $s$ quark  as a function of $\omega$ for several  values of $T$ with  $\mu=0$, here the result of $T =0.01$ GeV is divided by 10.}
\label{susceptibility.pdf}
\end{figure}
\end{widetext}

In order to have a better understanding for the spin polarization of  quark with rotation, we plot the second-order spin polarizations of the rotating system as a function of $\omega$ for several temperatures with  the chemical potential equals zero in Fig. \ref{p2polarization.pdf}. In the case of the temperature is very low, the second-order spin polarization begins to occur at a certain value of the  angular velocity and increasing with the angular velocity increases, until it reaches the highest value, when the angular velocity is increased further the second-order spin polarization becomes  weaker and finally disappears, the reason why the second-order  quark spin polarization disappears at large angular velocity is that when the angular velocity becomes larger, the dynamical mass of the quark becomes smaller and the chiral symmetry is gradually restored so that the  second-order quark spin polarization does not occur.

On the other hand, in order to understand the properties of the quark matter under the rotation system better,  it is helpful to study the behavior  of baryon number susceptibility. We now turn to  study the QNS and take into account the influence of angular velocity on  QNS.
Firstly, we consider the  case of zero chemical potential, in Fig. \ref{susceptibility.pdf}  we  plot the susceptibilities of $u$ and  $s$ quark as a function of angular velocity for several fixed values of temperatures with zero chemical potential, to be clear, the result of $T =0.01$ GeV is divided by 10.  It is evident from the figure that the susceptibilities of $u$ and $s$ quark increase as the $\omega$ increases when  $\omega$  is smaller than a certain value for all the temperatures, while the susceptibilities decrease
as the $\omega$ increases when  $\omega$  is larger than another certain value. And it also can be seen from the figure that the temperature and the angular velocity  play an important role in the susceptibility, at low temperature the quark chiral symmetry is broken spontaneously, however with the increasing of the angular velocity the chiral symmetry is restored, so  we can see at low temperature and low angular velocity the susceptibility is very small and  at low temperature and large angular velocity  the susceptibility disappears, however if the temperature is high the susceptibility always occurs and the angular velocity plays a slightly effect on the rotating matter.

It is fairly clear  that  the  susceptibility  is a finite quantity that is furthermore sensitive to the mass of the quark, so next we will discuss the differences of the $u$ and $s$ quark number susceptibilities under the rotation. When angular velocity is small and temperature is high we find that the susceptibility of $u$ is larger than the susceptibility of $s$ and when angular velocity and temperature are both small we find that it is easy for the susceptibility of $u$ quark to occur, for instance, at $T=0.01$ GeV, the $u$  quark number susceptibility starts to occur at $\omega=0.5$  GeV while for $s$   quark number susceptibility starts to occur at $\omega=0.9$ GeV.  As we can see from the figure  that the curves have  the highest values   only for some certain regions of the angular velocity, and at low temperature we  find that there exist a narrower region obvious  changes  the QNS  when comparing that of  $s$ quark  with that of  $u$ quark at $T=0.01$ GeV,  it is clearly can be seen from the curves that QNS changes little when the angular velocity is changed below 0.9 GeV or above 1.5  GeV for the $s$ quark, however for that of $u$ quark  the angular velocity is  below 0.5  GeV or above 1.5  GeV.   It is very clear from the figure that the contribution from the angular velocity becomes dominant when $\omega\geq1.1$ and the peaks of the susceptibilities  appear at almost the same angular velocity. At all  values of the chemical potential the behaviors of the $u$ quark and $s$ quark number susceptibility are very similar with increasing the angular velocity reveal that when the angular velocity is large the role played by the mass of different quarks becoming weaker and weaker and finally almost can be ignored.

\begin{widetext}

\begin{figure}[!htbp]
\setlength{\unitlength}{1mm}
\centering
\includegraphics[height=8cm]{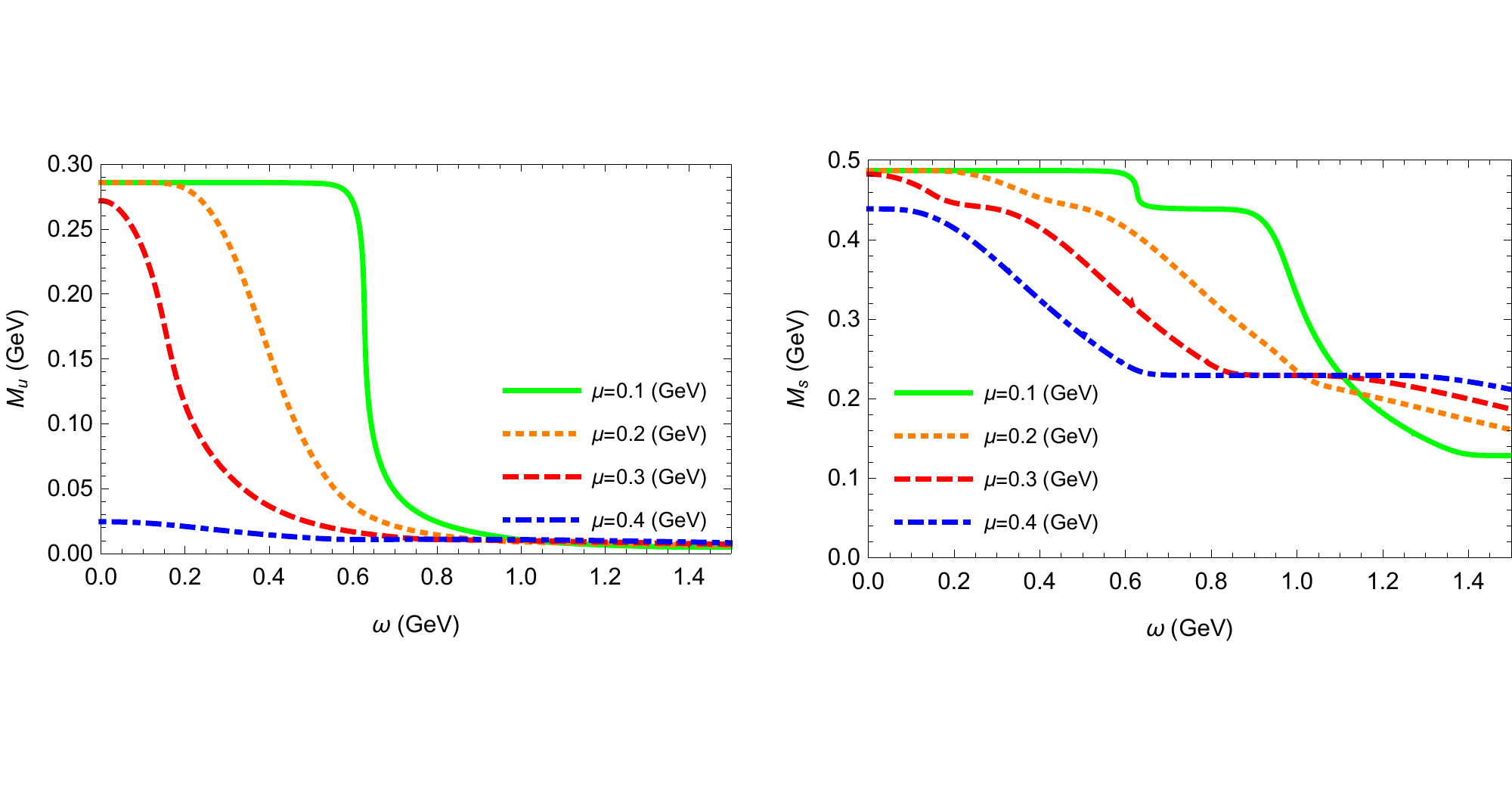}
\caption[]{(Color online) The mean field mass gap of $u$ and $s$ quark  as a function of $\omega$ for several fixed values of $\mu$ at $T=0.01$ GeV.}
\label{gapequation.with.mu.pdf}
\end{figure}

\end{widetext}

\begin{figure}[!htbp]
\setlength{\unitlength}{1mm}
\centering
\includegraphics[height=5cm]{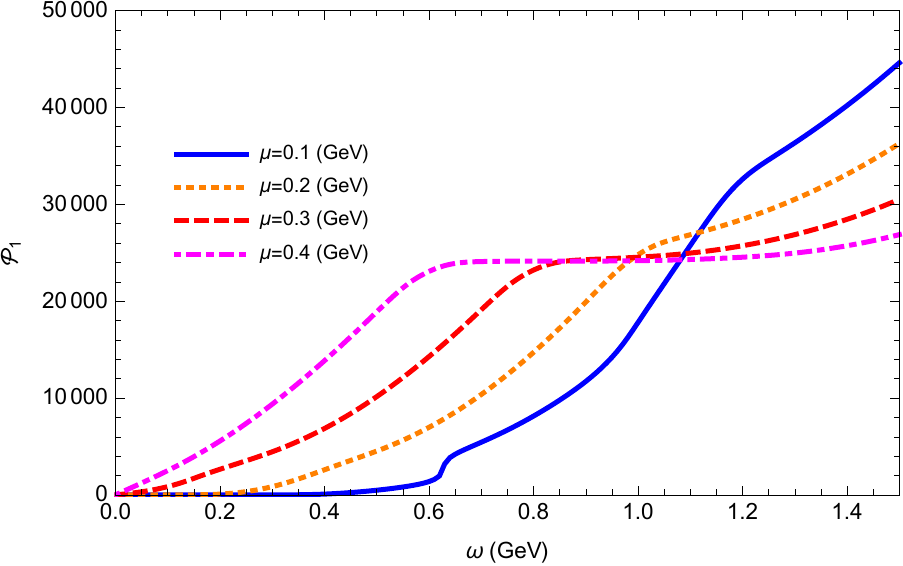}
\caption[]{(Color online) First-order spin polarizations  of the rotating system   as a function of $\omega$ for several fixed nonzero values of $\mu$ at $T=0.01$ GeV.}
\label{p1polarizationwithmu.pdf}
\end{figure}
Let us now discuss the  behavior of mean field mass gap of the quark  at very low temperature with nonzero chemical potential, in Fig. \ref{gapequation.with.mu.pdf}  we plot $M_{u}$ and $M_{s}$ as a function of $\omega$ for a variety of values of $\mu$ at $T=0.01$ GeV, respectively.
Comparing with the Fig. \ref{gapmu0.pdf}, it is clear that a nonzero value of the chemical potential affects the first-order phase transition, at $T=0.01$ GeV,  there does not exist a sudden drop for the mean field mass gap when $\mu$ is large both for $u$ and $s$ quark, which indicating there exists  a  suppression effect for the chemical potential  on the  phase transition. At large chemical potential the chiral condensate vanishes with increasing $\omega$ via a smooth crossover. From the figure we can also see that there is a different  behavior between  $u$ and $s$ quark,  the $u$ quark  is more affected by the presence of the chemical potential and angular velocity than $s$ quark because the $s$ quark has a substantial mass even after the chiral phase transition.

In   Fig. \ref{p1polarizationwithmu.pdf}, the plots of the first-order spin polarization  of the rotating system as a function of angular velocity at $T=0.01$ GeV are presented for nonzero chemical potential $\mu=0.1, 0.2, 0.3, 0.4$ GeV, respectively. As is clearly evident from the figure, the first-order spin polarizations  of the rotating system increase with increasing angular velocity. And it can be seen that the first-order spin polarizations are affected obviously by the  nonzero quark chemical potential.
With increasing angular velocity, the first-order spin polarization  of the system will first take place at larger quark chemical potential, for example, the first-order spin polarization  of $u$ start to occur aound $\omega=0.2$  GeV and $\omega=0.4$ GeV for $\mu=0.2$ GeV and $\mu=0.1$ GeV, respectively.  Fig. \ref{p1polarizationwithmu.pdf} also demonstrates that at very large quark chemical potential the first-order spin polarization  of the system first  quickly reach a certain value then is only relatively slowly varying with angular velocity,
this can be understood by noting that at large  value of quark chemical potential the chiral symmetry restored quickly.

\begin{figure}[!htbp]
\setlength{\unitlength}{1mm}
\centering
\includegraphics[height=5cm]{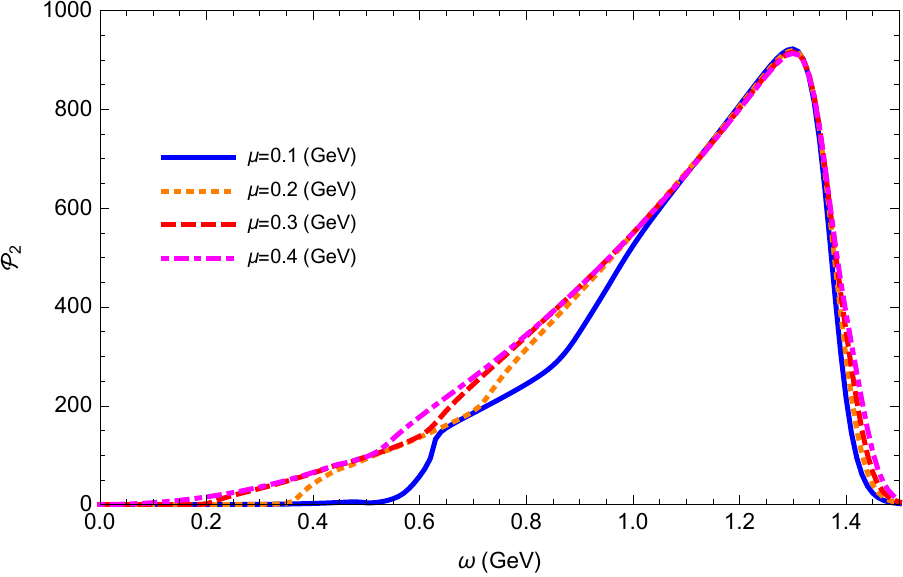}
\caption[]{(Color online) Second-order spin polarizations  of the rotating system as a function of $\omega$ for several fixed nonzero values of $\mu$ at $T=0.01$ GeV.}
\label{p2polarizationwithmu.pdf}
\end{figure}

Fig. \ref{p2polarizationwithmu.pdf} displays the results of the second-order  polarizations   of the system   as a function of $\omega$ for several fixed nonzero values of chemical potential at $T=0.01$ GeV.  In particular, a non-monotonic is identified,
with the second-order  polarizations to first increase, reach a maximum,  and then decrease sharply.
This behaviour is the combined effect of the suppression in both the angular velocity itself and the chemical potential at very low temperature.

\begin{widetext}

\begin{figure}[!htbp]
\setlength{\unitlength}{1mm}
\centering
\includegraphics[height=8cm]{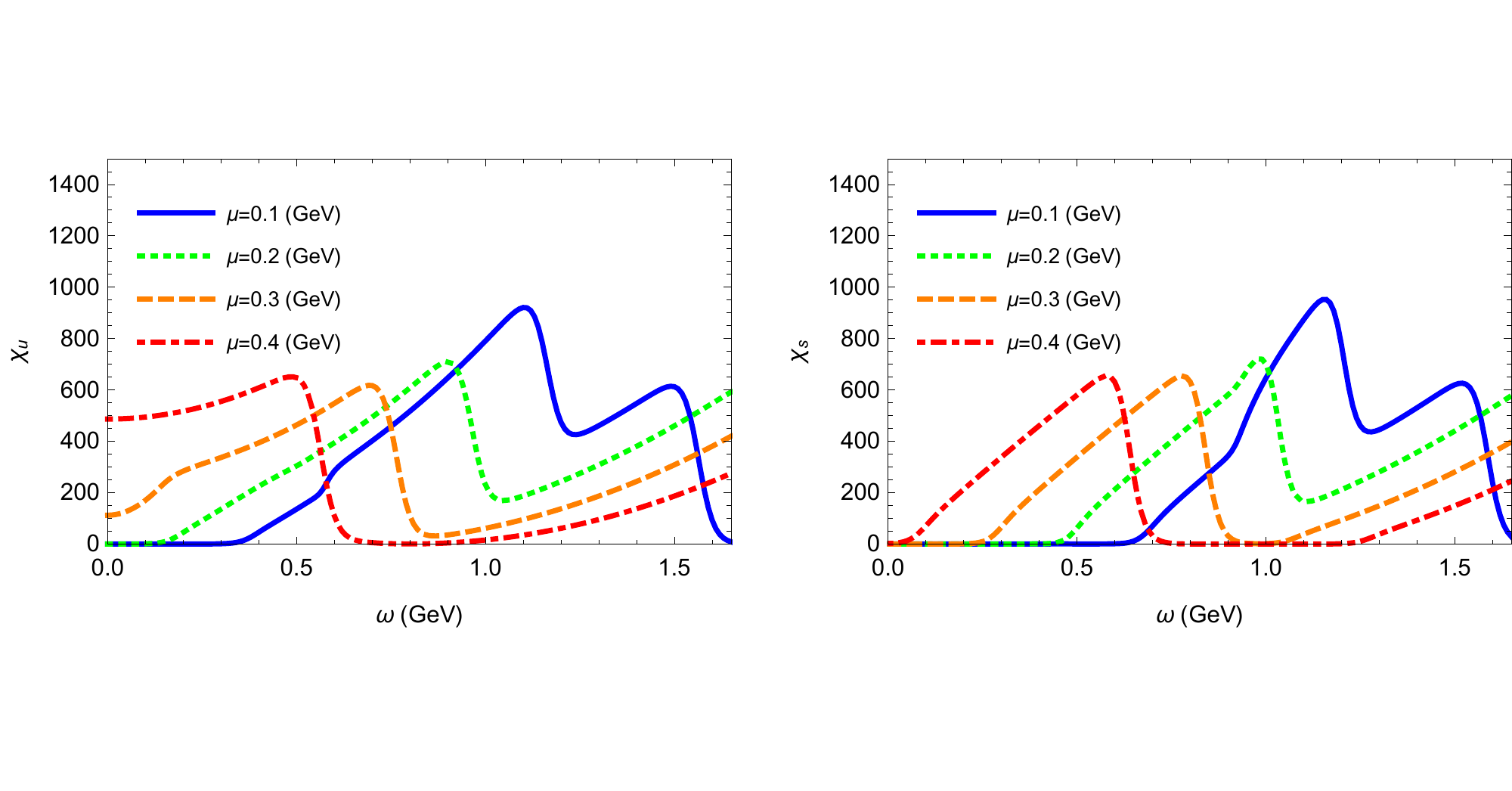}
\caption[]{(Color online) Susceptibilities of $u$ and $s$ quark  as a function of $\omega$ for several fixed nonzero values of $\mu$ at $T=0.01$ GeV.}
\label{susceptibilityt001mu.pdf}
\end{figure}
\end{widetext}

Next, we will analyze the  patterns of second-order susceptibility of quark with rotation at nonzero chemical potential. let us first consider the effect of angular velocity $\omega$ on second-order susceptibility of the quarks at very low temperature. We plot the second-order  susceptibilities of $u$ and $s$ quark   as a function of $\omega$ for several fixed nonzero values of chemical potential at $T=0.01$ GeV in Fig. \ref{susceptibilityt001mu.pdf}, respectively. From  the figure we find  the susceptibility   have similar behavior with respect to angular velocity. we can see if one  considers the case of nonzero chemical potential, the behavior of QNS changes considerably and is quite dependent on the angular velocity. As shown in the figure  one can see the dependence of QNS on angular velocity is complicated that the QNS increases with the increasing  $\omega$ when $\omega$ is smaller than a certain value while the QNS decreases with the increasing  $\omega$ when $\omega$ excesses  another certain value, which indicates that the rotation matter may provide some new and helpful results  to study the phase transition.  And there are some interesting changes compared with the situation of  zero chemical potential, at $T=0.01$ GeV the curves of   susceptibility of the quark have two peaks, which are very different compared with that in  Fig. \ref{susceptibility.pdf}, the curves of QNS have such behavior because the gap mass with nonzero chemical potential are different from that the case with zero chemical potential and for any angular velocity which should satisfy the gap equation, whose constraint will have an effect to the susceptibility. In addition, a plausible explanation for this phenomenon is that the rotational velocity serves as an effective chemical potential and exhibits a non-trivial  behavior such that the competition between the chemical potential and angular velocity renders the quark number susceptibility to reach its maximum in such manner.

\begin{widetext}

\begin{figure}[!htbp]
\setlength{\unitlength}{1mm}
\centering
\includegraphics[height=8cm]{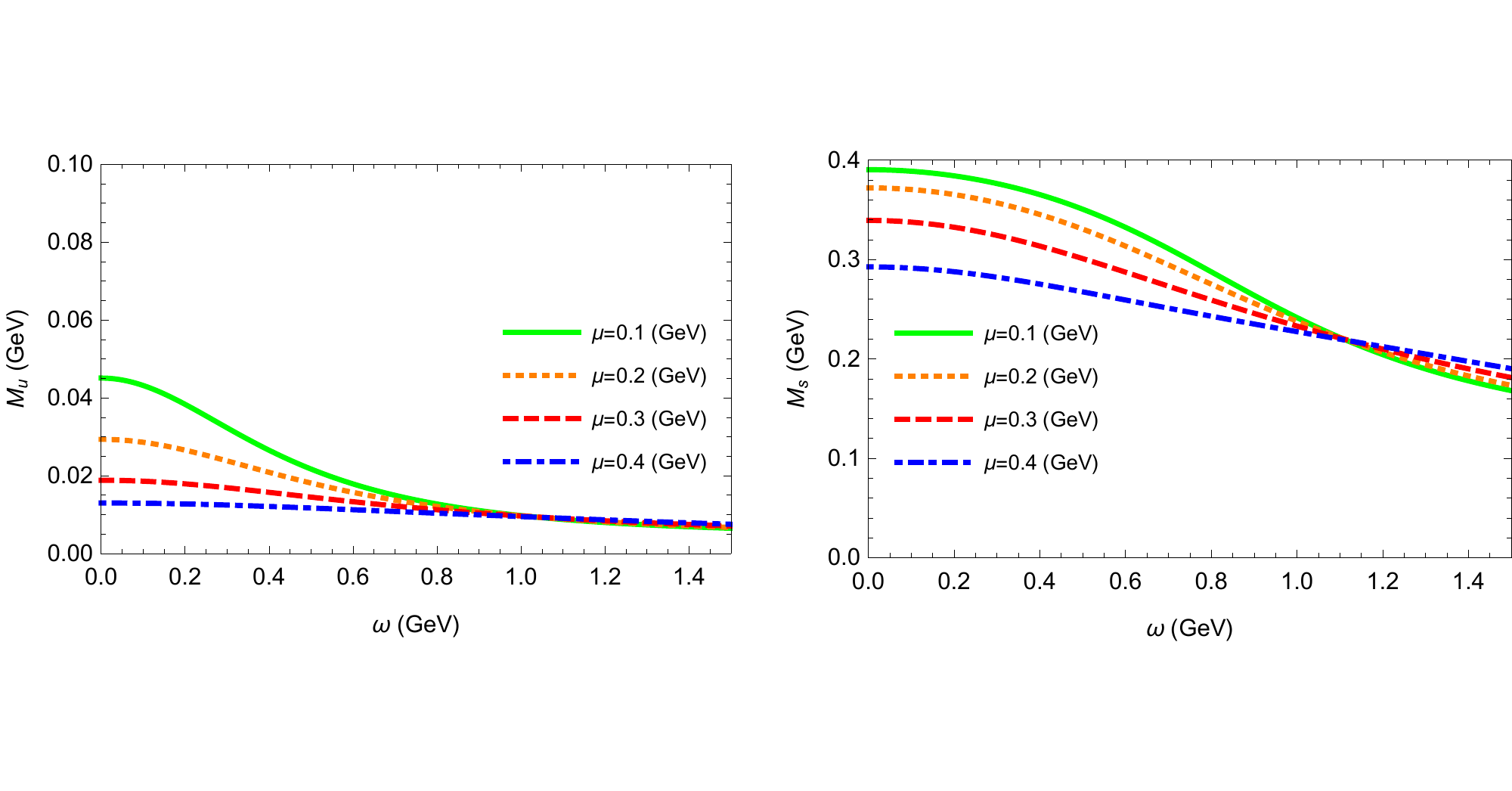}
\caption[]{(Color online) The mean field mass gap $M_{u}$ and $M_{s}$   as a function of $\omega$ for several fixed values of $\mu$ at $T=0.2$ GeV.}
\label{gapequation.witht020.mu.pdf}
\end{figure}

\begin{figure}[!htbp]
\setlength{\unitlength}{1mm}
\centering
\includegraphics[height=8cm]{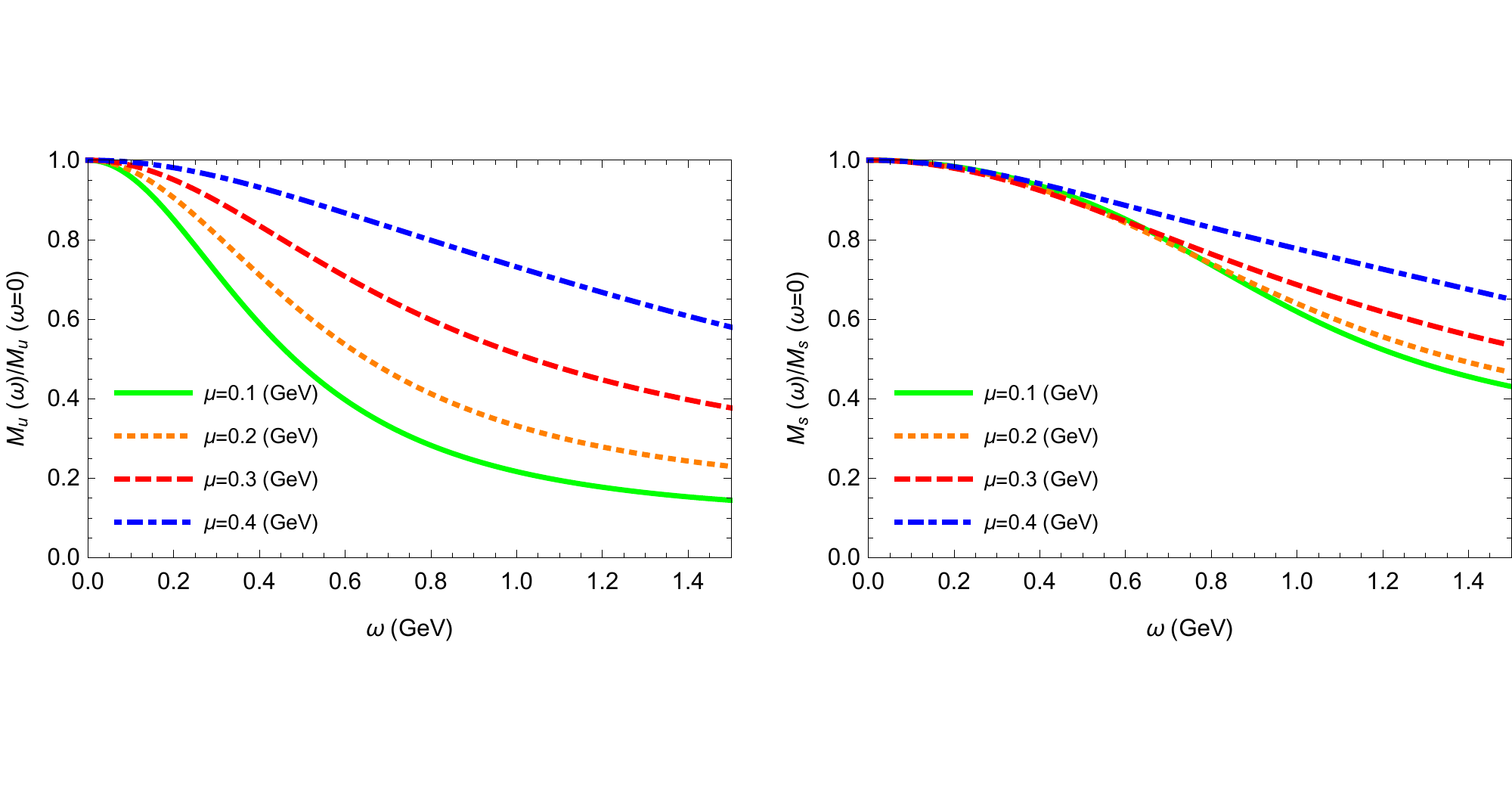}
\caption[]{(Color online) $M_{u}(\omega)/M_{u}(\omega=0)$ and $M_{s}(\omega)/M_{s}(\omega=0)$ as a function of $\omega$ with several nonzero chemical potentials at $T=0.2$ GeV.}
\label{gapequationratio.witht020.mu.pdf}
\end{figure}
\end{widetext}

It may need a bit more explanation about  the part played by the angular velocity $\omega$ with nonzero chemical potential at high temperature. Fig. \ref{gapequation.witht020.mu.pdf} shows the  mean field mass gap $M_{u}$  and $M_{s}$    versus $\omega$ with several fixed values of $\mu$ at $T=0.2$ GeV, obviously from the figure we can see there is a generally rotational suppression effect on the chiral condensate for the system at high temperature with nonzero chemical potential. In order to have better understanding  in Fig. \ref{gapequationratio.witht020.mu.pdf} we plot  $M_{u}(\omega)/M_{u}(\omega=0)$ and $M_{s}(\omega)/M_{s}(\omega=0)$ as a function of $\omega$ with several nonzero chemical potentials at $T=0.2$ GeV, respectively. We can see the influence of the angular velocity  to the light quark and strange quark is different.  The figure shows that $M_{u}$ is  much affected due to the current mass of the $u$ quark is very small and whose chiral symmetry can be easily  restored compared to that of $s$ quark.

\begin{figure}[!htbp]
\setlength{\unitlength}{1mm}
\centering
\includegraphics[height=5cm]{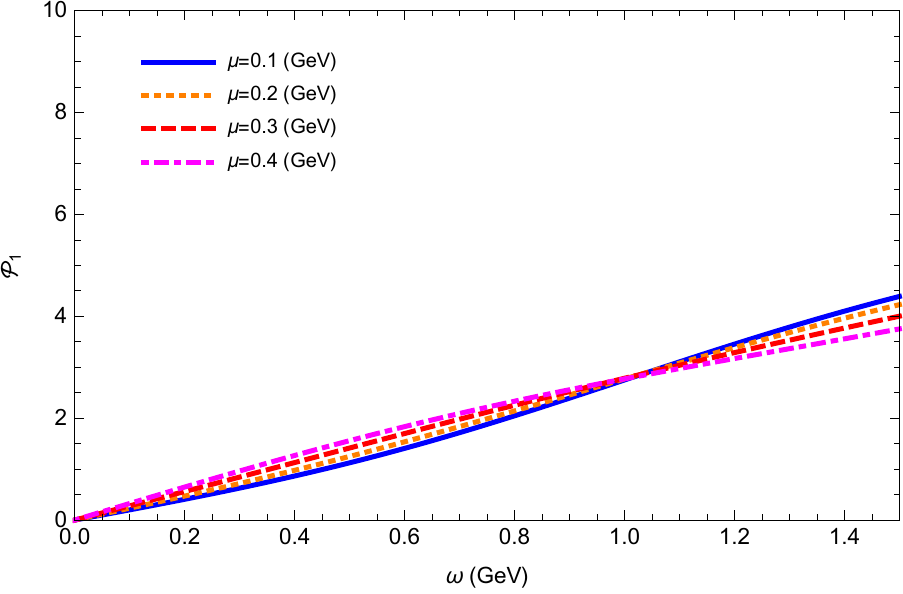}
\caption[]{(Color online)  First-order spin polarizations  of the rotating system as a function of $\omega$ for several fixed values of $\mu$ at $T=0.2$ GeV.}
\label{p1polarizationt020mu.pdf}
\end{figure}
\begin{figure}[!htbp]
\setlength{\unitlength}{1mm}
\centering
\includegraphics[height=5cm]{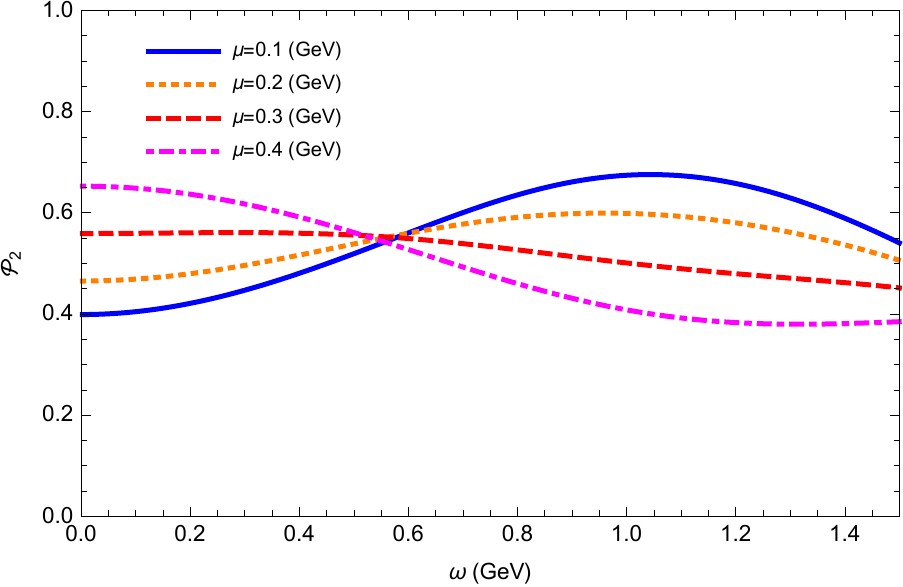}
\caption[]{(Color online) Second-order spin polarizations  of the rotating system  as a function of $\omega$ for several fixed nonzero values of $\mu$ at $T=0.2$ GeV.}
\label{p2polarizationt020mu.pdf}
\end{figure}
In Fig. \ref{p1polarizationt020mu.pdf}, we plot the first-order spin polarization  of the system  as a function of $\omega$ for several fixed values of $\mu$ with $T=0.2$ GeV. From Fig. \ref{p1polarizationt020mu.pdf} we find that at this temperature the first-order quark spin polarization of the system always occurs with increasing the angular velocity for all the nonzero chemical potential, at large chemical potential ($\mu=0.4$ GeV), the first-order spin polarization  of the system increase almost linearly with increasing the angular velocity.  We now consider the effects of angular velocity $\omega$ on second-order spin polarization the system at $T=0.2$ GeV with chemical potential $\mu=0.1, 0.2, 0.3, 0.4$ GeV, respectively. From the Fig. \ref{p2polarizationt020mu.pdf} one immediately makes the following observations,  in the case of  high temperature the variation of second-order spin polarization of the system with  angular velocity is complicated, we find that at small chemical potential  the second-order spin polarization of the system have a peak  structure with the increase of angular velocity,  when $\omega<1.0$ GeV at small chemical potential ($\mu=0.1$ GeV) the second-order spin polarization  of the system increases with increase in angular velocity, while at large chemical potential ($\mu=0.4$ GeV) that decreases with increase in angular velocity.

\begin{widetext}

\begin{figure}[!htbp]
\setlength{\unitlength}{1mm}
\centering
\includegraphics[height=8cm]{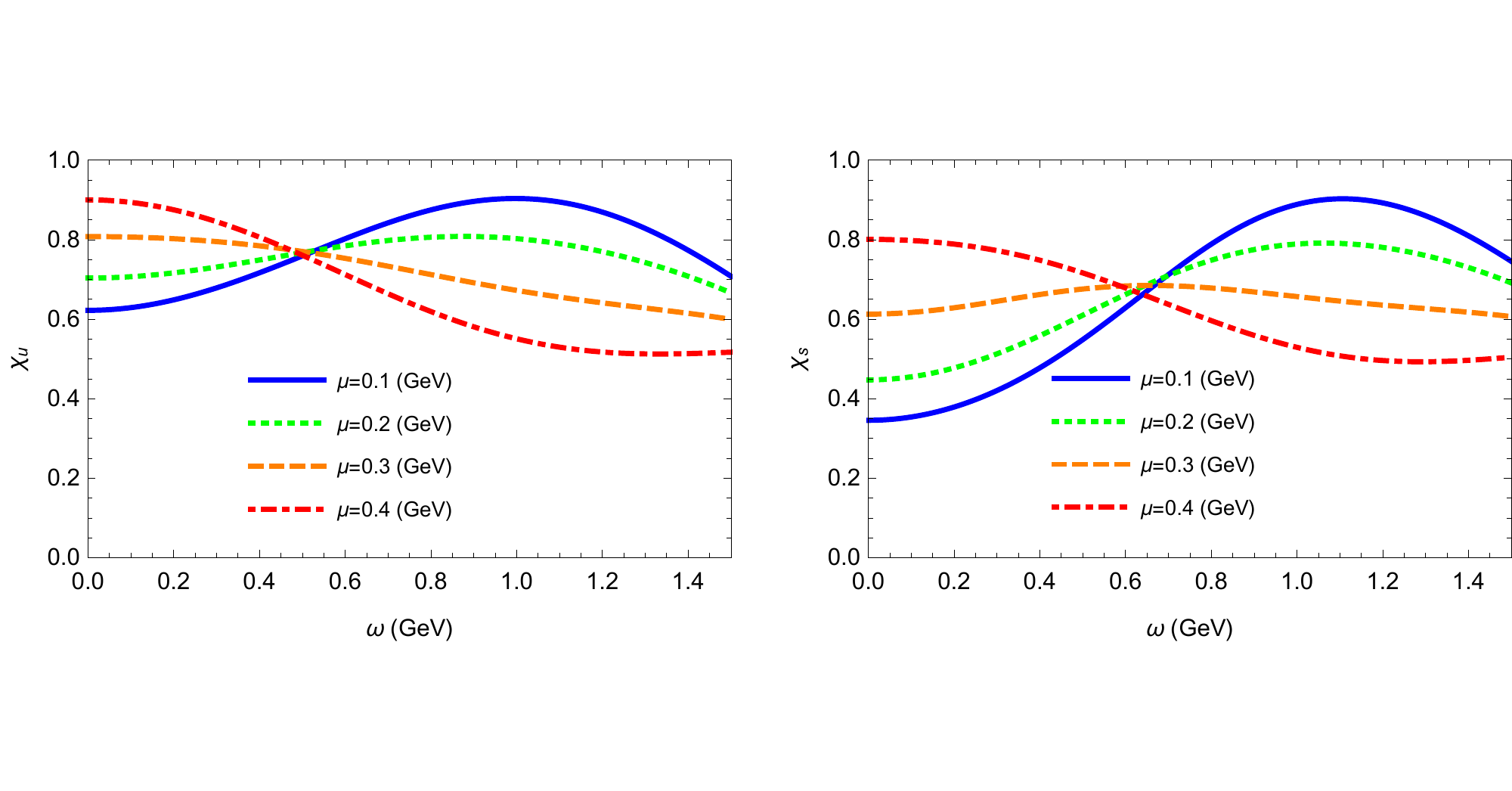}
\caption[]{(Color online) Second-order   susceptibilities  of $u$ and $s$ quark   as a function of $\omega$ for several fixed nonzero values of $\mu$ at $T=0.2$ GeV.}
\label{susceptibility.witht020.mu.pdf}
\end{figure}
\end{widetext}

Next, we would like to  consider the effects of angular velocity $\omega$ on second-order susceptibility of the quarks at high temperature with  several fixed nonzero values of chemical potential and the numerical results are shown in Fig. \ref{susceptibility.witht020.mu.pdf}.
It can be  found that at such high temperature   the second-order  susceptibility of the quarks can always occur, although  the maximum value of susceptibility is very small compared to the situation in  the Fig. \ref{susceptibilityt001mu.pdf}, which means that the contribution of angular velocity to the susceptibility is suppressed by high temperature. From Fig. \ref{susceptibility.witht020.mu.pdf} one could infer the dependence of the second-order susceptibility of the quarks on the quark current mass when $\omega<0.5$ GeV, the values of the susceptibility of the strange quark is  smaller compared to that of light quarks for the same quark chemical potential. However, with the increase of $\omega$, the behavior of the both quarks is very similar which means that at high temperature and large chemical potential the large angular velocity takes the predominant role compared to different current mass of the quarks. From  Fig. \ref{susceptibility.witht020.mu.pdf} one could also infer the dependence of susceptibility on the chemical potential at high temperature, at small $\omega$, the susceptibility increases with the increasing $\mu$. at large $\omega$, the susceptibility decreases with the increasing $\mu$.

Let us move on to the topic of  QCD phase diagram. The investigation of the phase diagram of QCD has been an active subject for many years.  There has been much progress on the study of the QCD phase diagram with lattice QCD (LQCD) simulations,  however,  at large  chemical potential the predictions made by the LQCD are not very reliable due to the sign problem of lattice gauge theory \cite{P.de Forcrand}. So, in order to investigate the QCD phase diagram many effective models have been proposed such as Nambu and Jona-Lasinio (NJL) models, quark-meson (QM) models, holographic QCD models, functional renormalization group (fRG),   Dyson Schwinger equations (DSE) as well as some extending modes of these \cite{S. P. Klevansky,H. Kohyama,C. D.  Roberts,R.  Alkofer,C. S. Fischer,I. C.,T.M. Schwarz,P. Zhuang,J.-W. Chen,J.-W. Chen1,W. Fan,W. Fan1,W.-J. Fu,E.S. Bowman,H. Mao,B.J. Schaefer,B.-J. Schaefer1,S.-X. Qin,J. Luecker,W.-J. Fu1,M. Buballa,Luis A. H. Mamani}. It is believed that the addition of external
influences or new parameter ranges yield an increasing number of interesting
results to the phase diagram \cite{J.N. Guenther}.

\begin{figure}[!htbp]
\setlength{\unitlength}{1mm}
\centering
\includegraphics[height=5cm]{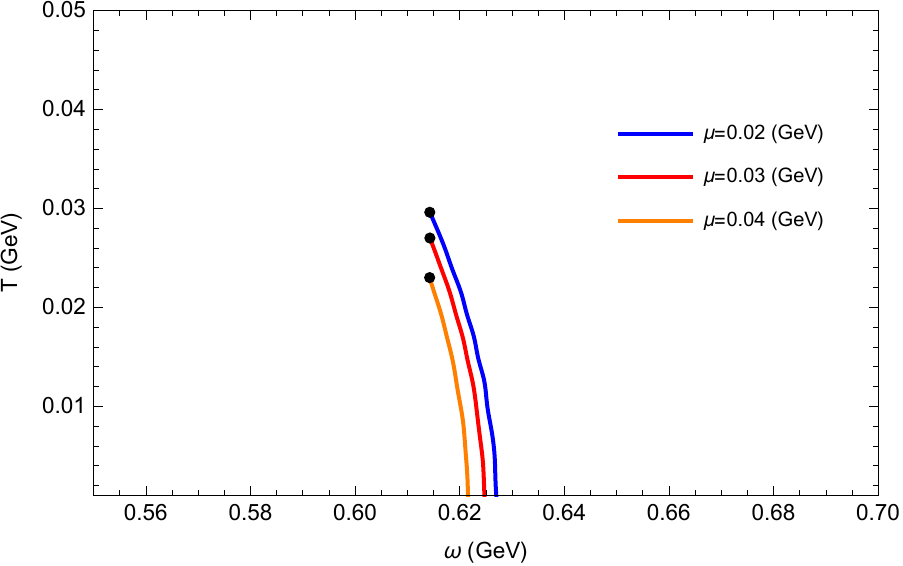}
\caption[]{(Color online) The first order phase transitions of  the system  on the $T$-$\omega$ plane.}
\label{phasetomegamu.pdf}
\end{figure}
And investigating the QCD matter under rotation is  a fascinating topic, apart from the chiral condensation, spin polarization and  number susceptibility of these matter, it is also of significant interest to explore  the effects of rotation on the phase transitions.
We now first explore the phase diagram in the $T$-$\omega$ plane, it is obvious that from the Fig. \ref{gapmu0.pdf} there exist first order phase transitions for the rotating system. Here  our main focus is on the first order phase transition  and its associated critical end point with rotation in the  presence of small chemical potential, in Fig.  \ref{phasetomegamu.pdf} we
show the first order phase diagram of the system in the $T$-$\omega$ plane with different chemical potentials  $\mu=0.02, 0.03$ and  $0.04 $ GeV, respectively.  We observe that the appearance of the first order phase transition line starting at $T\sim 0$ GeV and terminating at the critical end point identified with a full dot and these first order phase transition lines move toward a higher temperature for decreasing $\omega$.  It is also found that the different chemical potentials change the boundary of phase diagram, and that a larger chemical potential shifts down the critical temperature. It is no doubt that there are many other discussions need to be done for the phase transition in the rotation system, here is just a beginning and this particular topic will be discussed in detail in future work.

\section{Conclusions and outlook\label{sec4}}
Finally, we want to summarize our results and give a brief outlook.
In this paper,
we have presented detailed analytic formulae for the quark matter under rotation  in three-flavor NJL model and
related topics have been investigated. In order to have a better understanding of the rotating system with finite density we have also introduced the chemical potential. We  studied the quark fields in cylindrical coordinates as well as
investigated the effect of the rotating  on the quark chiral condensate, quark spin polarization and quark number susceptibility at finite temperature with or without finite chemical potential in this model.  We found that the angular velocity  plays a very crucial role in these topics, at low temperature, small chemical potential and small angular velocity the chiral symmetry is broken spontaneously, while at  large enough angular velocity the chiral symmetry is restored. Our numerical analysis shows that the rotation suppresses the  chiral condensation and enhances  the first-order quark spin polarization, however for the second-order quark spin polarization and quark number susceptibility the effect is  very complicated, which can be found to have a peak structure.

We have also explicitly computed these quantities in the rotation system in the presence of  chemical potential, we found  that  the nonzero chemical potential affects  and makes the chiral condensate, quark spin polarization and quark number susceptibility have different behaviors. At very low temperature the chiral condensate experiences a first order transition when $\omega$ exceeds a certain value with zero chemical potential, while at larger chemical potential the first order transition is suppressed  and changed to a crossover transition. It can be also observed that at very low temperature the quark number susceptibilities have two maxima, a plausible explanation for this phenomenon is that the rotational velocity serves as an effective chemical potential and exhibits a non-trivial  behavior such that the competition between the chemical potential and angular velocity renders the quark number susceptibility to reach its maximum in such manner. In this paper we  especially considered the contributions from $s$ quark and made some comparisons between $u$ quark and $s$ quark and found that at small angular velocity the part played by current mass to these phenomena is important, however, at sufficiently large angular velocity, the contributions played by different quarks to these phenomena are almost equal. In addition, we  explored the phase diagram in the $T$-$\omega$ plane,  we observe that there exist first order phase transitions for the rotating system and  the first order phase transition lines move toward a
higher temperature for decreasing angular velocity. It is also found that the different chemical potentials change the boundary of phase diagram, and that a larger chemical potential shifts down the critical temperature.  Based on the interpretations made above, it would be possible to judge and forecast these  phenomena of quark matter under rotation if we jointly take angular velocity, chemical potential and temperature factors into consideration  in the three-flavor NJL model. We expect these studies to play an
important role in help understanding the properties of strongly interacting rotating matter.

The theoretical interest in relativistic rotating systems is being revived in many different physical environments, which is undoubtedly calls for more investigation. For instance, the related effects of rotating fermions inside a cylindrical boundary \cite{Victor E. Ambrus},  the investigation of a possible phase structure under rotation including the $s$ quark, especially the  exploration to those regions of the phase diagram that cannot be reached on the lattice yet. The NJL model describes only quarks and antiquarks and neglects the gluons, so it is also very worth to extend the rotation system to the Polyakov extended Nambu and Jona-Lasinio (PNJL) model \cite{C. Ratti}, which will consider the  complex interactions between quarks and gluons and that the chiral symmetry restoration as well as the effect of quark confinement in PNJL under rotation  may provide needed insight into the QCD. This model will have a clearer picture considering the constraint from the experimental related to rotation conducted at numerous research facilities worldwide as the Brookhaven National Laboratory (BNL), the European Organization for Nuclear Research (CERN) and the GSI Helmholtz Centre for Heavy Ion Research (GSI), due to one can see that some limits can be made for the parameter space under the given assumptions. It would be interesting to use the results obtained in this paper  to investigate  these topics discussed here, and we leave these as our  further study.\\

\section*{Acknowledgements}
We specially thank Jinfeng Liao for the early involvement in the work and enlightening discussions, and thank Hui Zhang, Akira Watanabe for discussions, corrections and comments. We also thank Shengqin Feng and Yafei Shi for their encouragement and discussions. The work has been supported by the National Natural Science Foundation of China (NSFC)
under Grant No.~11647174, No.~11875178, No.~11801311, the Key Laboratory of Quark and Lepton Physics Contracts under Grant No.~QLPL201905 and the  Science  Research  Foundation of China Three Gorges University under Grant No.~KJ2015A007. AH acknowledges support by the NSF Grant No.~PHY1913729 and by the U.S. Department of Energy, Office of Science, Office of Nuclear Physics, within the framework of the Beam Energy Scan Theory (BEST) Topical Collaboration. AH is also grateful to the Fundamental Research Funds for the Central Universities.

\begin{appendix}
\section{The brief description of fermions under  rotation \label{A}}

The properties of fermions under global rotation are relevant to a number of problems as discussed above, so it is important to choose  an appropriate complete set of commuting operators in the cylindrical coordinates. In this section, we will start  from the Dirac equation in the rotating frame, then we will derive the eigenvectors of the those complete set of commuting operators.

The general Lagrangian of the rotating fermions is written in the following way \cite{A. L. Fetter, Y. Jiang1, A. Yamamoto}
\begin{widetext}
\begin{eqnarray}
{\cal L} = \bar \psi \left[ {i\gamma ^\mu  \partial _\mu   - m + \left( {\gamma ^0 } \right)^{ - 1} \left( {\left( {\mathord{\buildrel{\lower3pt\hbox{$\scriptscriptstyle\rightharpoonup$}}
\over \omega }  \times \mathord{\buildrel{\lower3pt\hbox{$\scriptscriptstyle\rightharpoonup$}}
\over x} } \right).\left( { - i\mathord{\buildrel{\lower3pt\hbox{$\scriptscriptstyle\rightharpoonup$}}
\over \partial } } \right) + \mathord{\buildrel{\lower3pt\hbox{$\scriptscriptstyle\rightharpoonup$}}
\over \omega } .\mathord{\buildrel{\lower3pt\hbox{$\scriptscriptstyle\rightharpoonup$}}
\over S} _{4 \times 4} } \right)} \right]\psi, \label{Lagrangian}
\end{eqnarray}
\end{widetext}
where $\psi$ is the quark field, $\omega$ is the angular velocity and $m$ is the  bare quark mass matrix, as a result of rotation, we can see the Dirac operator includes the orbit-rotation coupling term and the spin-rotation coupling term, and we have defined
\begin{eqnarray}
\mathord{\buildrel{\lower3pt\hbox{$\scriptscriptstyle\rightharpoonup$}}
\over S} _{4 \times 4}  = \frac{1}{2}\left( {\begin{array}{*{20}c}
   {\mathord{\buildrel{\lower3pt\hbox{$\scriptscriptstyle\rightharpoonup$}}
\over \sigma } } & 0  \\
   0 & {\mathord{\buildrel{\lower3pt\hbox{$\scriptscriptstyle\rightharpoonup$}}
\over \sigma } }  \\
\end{array}} \right),
\end{eqnarray}
whose z-component related to the spin polarization of the quark. The corresponding Hamiltonian of Eq. (\ref{Lagrangian}) in momentum space reads
\begin{widetext}
\begin{eqnarray}
\hat {\cal H} = \gamma ^0 \left( {\mathord{\buildrel{\lower3pt\hbox{$\scriptscriptstyle\rightharpoonup$}}
\over \gamma } .\hat {\vec p} + m} \right) - \mathord{\buildrel{\lower3pt\hbox{$\scriptscriptstyle\rightharpoonup$}}
\over \omega } .\left( {\mathord{\buildrel{\lower3pt\hbox{$\scriptscriptstyle\rightharpoonup$}}
\over x}  \times \hat {\vec p} + \mathord{\buildrel{\lower3pt\hbox{$\scriptscriptstyle\rightharpoonup$}}
\over S} _{4 \times 4} } \right) = \hat {\cal H}_0  + \mathord{\buildrel{\lower3pt\hbox{$\scriptscriptstyle\rightharpoonup$}}
\over \omega } .\hat {\vec J},
\end{eqnarray}
\end{widetext}
here
\begin{eqnarray}
\hat {\vec J} = \mathord{\buildrel{\lower3pt\hbox{$\scriptscriptstyle\rightharpoonup$}}
\over x}  \times \hat {\vec p} + \mathord{\buildrel{\lower3pt\hbox{$\scriptscriptstyle\rightharpoonup$}}
\over S} _{4 \times 4},
\end{eqnarray}
and the first term is the contribution of the angular momentum, the second term is the contribution of the spin angular momentum.

Now considering the energy eigenvalue equation
\begin{eqnarray}
\hat {\cal H}\psi  = E\psi, \label{eigenvalue}
\end{eqnarray}
here $\psi$ is the four-component spinors and  can be written in terms of two-component spinors as
\begin{eqnarray}
\psi  = \left( {\begin{array}{*{20}c}
   \phi   \\
   \chi   \\
\end{array}} \right),
\end{eqnarray}
substituting the Hamiltonian above to the energy eigenvalue equation, then Eq. (\ref{eigenvalue}) transforms simply as
\begin{eqnarray}
\left\{ \begin{array}{l}
 \left( {E - m + \omega _z J_z } \right)\phi  = \mathord{\buildrel{\lower3pt\hbox{$\scriptscriptstyle\rightharpoonup$}}
\over \sigma } . {\hat {\vec{p} }}\chi  \\
 \left( {E + m + \omega _z J_z } \right)\chi   = \mathord{\buildrel{\lower3pt\hbox{$\scriptscriptstyle\rightharpoonup$}}
\over \sigma } . {\hat {\vec{p} }}\phi  \\
 \end{array} \right.
\end{eqnarray}
here,
\begin{eqnarray}
J_z  = L_z  + \frac{1}{2}\sum _z,
 \end{eqnarray}
which is the  z-component of total angular momentum, then, we consider the $z$-component angular momentum eigenvalue equation

\begin{eqnarray}
J_z \psi {\rm{ = }}\left( {n + \frac{1}{2}} \right)\psi,\label{angular momentum}
\end{eqnarray}
after some derivations, we can get the following equation
\begin{widetext}
\begin{eqnarray}
\left( {E - m + \omega _z \left( {n + \frac{1}{2}} \right)} \right)\left( {E + m + \omega _z \left( {n + \frac{1}{2}} \right)} \right)\phi  = \left(\mathord{\buildrel{\lower3pt\hbox{$\scriptscriptstyle\rightharpoonup$}}
\over \sigma } . {\hat {\vec{p} }} \right)^2 \phi,\label{equation}
\end{eqnarray}
\end{widetext}
it is very convenient to make the transform Cartesian coordinate to the Cylindrical coordinate, here separation variable method is applied and $\phi$ takes the form of
\begin{eqnarray}
\phi = f\left( \theta  \right)g\left( r \right)h\left( z \right),
\end{eqnarray}
and the solution for two-component spinors $\varphi$ in Eqs. (\ref{angular momentum}) has the form
\begin{eqnarray}
 f(\theta)= \left( {\begin{array}{*{20}c}
   {e^{in\theta } }  \\
   {e^{i\left( {n + 1} \right)\theta } }  \\
\end{array}} \right),\label{spinors}
\end{eqnarray}
substituting Eq. (\ref{spinors}) into the Eq. (\ref{equation}) after some tedious calculations, we find that $g\left( r \right)$ satisfies the Bessel-type equation as follows,
\begin{widetext}
\begin{eqnarray}
r^2 \frac{{\partial ^2 g\left( r \right)}}{{\partial r^2 }} + r\frac{{\partial g\left( r \right)}}{{\partial r}} + \left( {r^2 p_t^2  - n^2 } \right)g\left( r \right) = 0,
\label{Bessel1}
\end{eqnarray}

\begin{eqnarray}
r^2 \frac{{\partial ^2 g\left( r \right)}}{{\partial r^2 }} + r\frac{{\partial g\left( r \right)}}{{\partial r}} + \left( {r^2 p_t^2  - \left( {n + 1} \right)^2 } \right)g\left( r \right) = 0,
\label{Bessel2}
\end{eqnarray}
the solutions of Eqs. (\ref{Bessel1}) and (\ref{Bessel2}) have the following form, respectively,
\begin{eqnarray}
g\left( r \right) = J_n \left( {p_t r} \right), J_{n+1} \left( {p_t r} \right),
\end{eqnarray}
\end{widetext}
where  $J$ is the Bessel function. In order to commute with other operators we must define the helicity operator, the general helicity operator has the following form
\begin{widetext}
\begin{eqnarray}
h_t  = \gamma ^5 .\gamma ^3 \frac{{\mathord{\buildrel{\lower3pt\hbox{$\scriptscriptstyle\rightharpoonup$}}
\over \sum } .\mathord{\buildrel{\lower3pt\hbox{$\scriptscriptstyle\rightharpoonup$}}
\over p} _t }}{{\left| {\mathord{\buildrel{\lower3pt\hbox{$\scriptscriptstyle\rightharpoonup$}}
\over p} _t } \right|}} = \frac{1}{i{\left| {\mathord{\buildrel{\lower3pt\hbox{$\scriptscriptstyle\rightharpoonup$}}
\over p} _t } \right|}}\left( {\begin{array}{*{20}c}
   0 & { - P_ -  } & 0 & 0  \\
   {P_ +  } & 0 & 0 & 0  \\
   0 & 0 & 0 & {P_ -  }  \\
   0 & 0 & { - P_ +  } & 0  \\
\end{array}} \right),
\end{eqnarray}
\end{widetext}
here, $p_{t}$ is the transverse momentum, $P_ +   = \hat p_x  + i\hat p_y$,  $P_ -   = \hat p_x  - i\hat p_y$ and in Cylindrical coordinates they have such forms
\begin{eqnarray}
P_ +   = - ie^{i\theta } \left( {\frac{\partial }{{\partial r}} + i\frac{1}{r}\frac{\partial }{{\partial \theta }}} \right),
\end{eqnarray}
\begin{eqnarray}
P_ -   =  - ie^{ - i\theta } \left( {\frac{\partial }{{\partial r}} - i\frac{1}{r}\frac{\partial }{{\partial \theta }}} \right),
\end{eqnarray}
which like shift operators when act on the terms including angular momentum quantum number $
e^{in\theta } J_n (p_t r)$, $e^{i\left( {n + 1} \right)\theta } J_{n + 1} (p_t r)$, respectively,   they satisfy the following relationship
\begin{eqnarray}
P_ +  e^{in\theta } J_n (p_t r) = ip_t e^{i\left( {n + 1} \right)\theta } J_{n + 1} (p_t r),
\end{eqnarray}
\begin{eqnarray}
P_ -  e^{i\left( {n + 1} \right)\theta } J_{n + 1} (p_t r) = -ip_t  e^{in\theta } J_n (p_t r).
\end{eqnarray}

Reconsidering the transverse helicity equation and the generalized orthogonality relation
\begin{eqnarray}
h_t \psi  = s\psi,
\end{eqnarray}

\begin{eqnarray}
\sum\limits_{n =  - \infty }^\infty  {\psi ^\dag  \psi  = 1},
\end{eqnarray}
here, $s=\pm1$ represent the transverse helicity, the solutions of the positive energy eigenvalues  are obtained as follows
 \begin{eqnarray}
u = \frac{1}{{\sqrt 2 }}\left( {\begin{array}{*{20}c}
   {e^{ip_z z} e^{i{\rm{n}}\theta } J_n \left( {p_t r} \right)}  \\
   {se^{ip_z z} e^{i\left( {n + 1} \right)\theta } J_{n + 1} \left( {p_t r} \right)}  \\
   {e^{ip_z z} e^{i{\rm{n}}\theta } J_n \left( {p_t r} \right)}  \\
   {se^{ip_z z} e^{i\left( {n + 1} \right)\theta } J_{n + 1} \left( {p_t r} \right)}  \\
\end{array}} \right),
\end{eqnarray}
where, $u$ is a  four-component spinor which must satisfy the Dirac equation
\begin{eqnarray}
(i \gamma^{\mu}\partial_{\mu}-m)u=0,
\end{eqnarray}
substituting $u$ into the Dirac equation gives
\begin{eqnarray}
\left( {\begin{array}{*{20}c}
   {E - m} & { - \sigma .p}  \\
   {\sigma .p} & { - E - m}  \\
\end{array}} \right)\left( {\begin{array}{*{20}c}
   {c_A u_A }  \\
   {c_B u_B }  \\
\end{array}} \right) = 0,
\end{eqnarray}
here $u_A$, $u_B$ are two-component spinors and $c_A$, $c_B$ are normalization constants, after some calculations we get
\begin{widetext}
\begin{eqnarray}
c_B u_B  = c_A \left( {\begin{array}{*{20}c}
   {\frac{{\left( {p_z  - isp_t } \right)}}{{E + m}}e^{ip_z z} e^{i{\rm{n}}\theta } J_n \left( {p_t r} \right)}  \\
   {\frac{{\left( { - sp_z  + ip_t } \right)}}{{E + m}}e^{ip_z z} e^{i\left( {n + 1} \right)\theta } J_{n + 1} \left( {p_t r} \right)}  \\
\end{array}} \right),
\end{eqnarray}
\end{widetext}
imposing the generalized completeness relation
\begin{eqnarray}
\sum\limits_{n =  - \infty }^\infty  {u^\dag  u = 1},
\end{eqnarray}
these constant factors can be determined and  finally we obtain the positive energy particle solutions with positive and negative helicity in the Dirac representation, which take the following explicit form
\begin{widetext}
\begin{eqnarray}
u = \frac{1}{2}\sqrt {\frac{{E + m}}{E}} \left( {\begin{array}{*{20}c}
   {e^{ip_z z} e^{i{\rm{n}}\theta } J_n \left( {p_t r} \right)}  \\
   {se^{ip_z z} e^{i\left( {n + 1} \right)\theta } J_{n + 1} \left( {p_t r} \right)}  \\
   {\frac{{p_z  - isp_t }}{{E + m}}e^{ip_z z} e^{i{\rm{n}}\theta } J_n \left( {p_t r} \right)}  \\
   {\frac{{ - sp_z  + ip_t }}{{E + m}}e^{ip_z z} e^{i\left( {n + 1} \right)\theta } J_{n + 1} \left( {p_t r} \right)}  \\
\end{array}} \right),
\end{eqnarray}
in exactly the same way, the negative-energy antiparticle solutions  are listed below

\begin{eqnarray}
v = \frac{1}{2}\sqrt {\frac{{E + m}}{E}} \left( {\begin{array}{*{20}c}
   {\frac{{p_z  - isp_t }}{{E + m}}e^{ - ip_z z} e^{i{\rm{n}}\theta } J_n \left( {p_t r} \right)}  \\
   {\frac{{ - sp_z  + ip_t }}{{E + m}}e^{ - ip_z z} e^{i\left( {n + 1} \right)\theta } J_{n + 1} \left( {p_t r} \right)}  \\
   {e^{ - ip_z z} e^{i{\rm{n}}\theta } J_n \left( {p_t r} \right)}  \\
   { - se^{ - ip_z z} e^{i\left( {n + 1} \right)\theta } J_{n + 1} \left( {p_t r} \right)}  \\
\end{array}} \right).
\end{eqnarray}
\end{widetext}

\end{appendix}


\begin{thebibliography}{199}
\bibitem{D. Kharzeev}
D. Kharzeev and A. Zhitnitsky, Nucl. Phys. A797, 67 (2007).
\bibitem{D. T. Son}
D.T. Son and P. Surowka, Phys. Rev. Lett. 103, 191601 (2009).
\bibitem{D. E. Kharzeev}
D.E. Kharzeev and D. T. Son, Phys. Rev. Lett. 106, 062301 (2011).
\bibitem{Y. Jiang}
Y. Jiang, X. G. Huang, and J. Liao, Phys. Rev. D 92, 071501 (2015).
\bibitem{Kharzeev}
Kharzeev, D. E., Liao, J., Voloshin, S. A. $\&$ Wang, G. Chiral magnetic and vortical effects
in high-energy nuclear collisionsA status report. Prog. Part. Nucl. Phys. 88, 1-28 (2016).
1511.04050.
\bibitem{L. P. Csernai}
L.P. Csernai, V.K. Magas and D.J. Wang, Phys. Rev. C 87, no. 3, 034906 (2013).
\bibitem{F. Becattini3}
F. Becattini et al., Eur. Phys. J. C 75, no. 9, 406 (2015).
\bibitem{Y. Jiang2}
Y. Jiang, Z.W. Lin and J. Liao, Phys. Rev. C 94, no. 4,
044910 (2016) Erratum: [Phys. Rev. C 95, no. 4, 049904
(2017)].
\bibitem{S. Shi}
S. Shi, K. Li and J. Liao, Phys. Lett. B 788, 409 (2019).
\bibitem{W. T. Deng}
W.T. Deng and X.G. Huang, Phys. Rev. C 93, no. 6,
064907 (2016).
\bibitem{L. G. Pang}
L.G. Pang, H. Petersen, Q. Wang and X.N. Wang, Phys.
Rev. Lett. 117, no. 19, 192301 (2016).
\bibitem{L. Adamczyk}
L. Adamczyk et al. [STAR Collaboration], Nature 548,
62 (2017) doi:10.1038/nature23004 [arXiv:1701.06657
[nucl-ex]].
\bibitem{F. Becattini4} F. Becattini, I. Karpenko, M. Lisa, I. Upsal and
S. Voloshin, Phys. Rev. C 95, no. 5, 054902 (2017).
\bibitem{X. L. Xia} X.L. Xia, H.Li, Z.B. Tang and Q. Wang, Phys. Rev. C
98, 024905 (2018).
\bibitem{Hui Zhang}
Hui Zhang, Defu Hou and Jinfeng Liao, arXiv:1812.11787v3 [hep-ph].
\bibitem{A. L. Fetter}
A.L. Fetter, Rev. Mod. Phys. 81, 647 (2009).
doi:10.1103/Rev Mod Phys. 81. 647.
\bibitem{Urban}
Urban, M.,  Schuck, P. 2008, Phys. Rev. A , 78, 011601.
\bibitem{Iskin}
Iskin, M.,  Tiesinga, E. 2009, Phys. Rev. A , 79, 053621.
\bibitem{R. Takahashi}
R. Takahashi, et al, Nature Physics volume 12 (2016).
\bibitem{J. Gooth}
J. Gooth et al., Nature 547, 324 (2017).



\bibitem{A. Vilenkin1}
A. Vilenkin, Parity Violating Currents in Thermal Radiation, Phys. Lett. 80B (1978) 150.

\bibitem{A. Vilenkin2}
A. Vilenkin, Macroscopic Parity Violating Effects: Neutrino Fluxes From RotatingBlack
Holes And In Rotating Thermal Radiation, Phys. Rev. D 20 (1979) 1807 [INSPIRE].

\bibitem{A. Vilenkin3}
A. Vilenkin, Quantum Field Theory At Finite Temperature In A Rotating System, Phys. Rev.
D 21 (1980) 2260 [INSPIRE].

\bibitem{M. Kaminski}
M. Kaminski, C.F. Uhlemann, M. Bleicher and J. Schaffner-Bielich, Anomalous
hydrodynamics kicks neutron stars, Phys. Lett. B 760 (2016) 170 [arXiv:1410.3833]
[INSPIRE].

\bibitem{N. Yamamoto}
N. Yamamoto, Chiral transport of neutrinos in supernovae: Neutrino-induced
uid helicity and helical plasma instability, Phys. Rev. D 93 (2016) 065017 [arXiv:1511.00933] [INSPIRE].

\bibitem{E. Shaverin}
E. Shaverin and A. Yarom, An anomalous propulsion mechanism, arXiv:1411.5581
[INSPIRE].


\bibitem{A. L. Watts}
A. L. Watts et al., Rev. Mod. Phys. 88, no.
2, 021001 (2016) doi:10.1103/Rev Mod Phys. 88. 021001 [arXiv:1602.01081 [astro-ph.HE]].
\bibitem{I. A. Grenier} I. A. Grenier and A. K. Harding, Comptes Rendus Physique 16, 641 doi: 10.1016/j. crhy. 2015. 08. 013
[arXiv:1509.08823 [astro-ph.HE]].
\bibitem{E. Berti}
E. Berti, F. White, A. Maniopoulou and M. Bruni,
Mon. Not. Roy. Astron. Soc. 358, 923 (2005)
doi:10.1111/j. 1365-2966. 2005. 08812. x [gr-qc/0405146].

\bibitem{A. Yamamoto}
A. Yamamoto and Y. Hirono, Phys. Rev. Lett.
111, 081601 (2013) doi:10.1103/Phys. Rev. Lett. 111. 081601 [arXiv:1303.6292 [hep-lat]].
\bibitem{Z. T. Liang}
Z.T. Liang and X.N. Wang, "Globally Polarized Quark Gluon Plasma in Noncentral A+A Collisions," Phys.
Rev. Lett. 94, 102301 (2005), [Erratum: Phys. Rev. Lett.
96, 039901 (2006)].
\bibitem{Sergei A. Voloshin}
Sergei A. Voloshin, "Polarized secondary particles in unpolarized
high energy hadron-hadron collisions?" (2004),
arXiv:nucl-th/0410089 [nucl-th].
\bibitem{F. Becattini5}
F. Becattini, F. Piccinini, and J. Rizzo, "Angular momentum
conservation in heavy ion collisions at very high
energy," Phys. Rev. C 77, 024906 (2008).
\bibitem{Z.T. Liang}
Z.T. Liang and X. N. Wang, Phys. Rev. Lett. 94 (2005) 102301, nuclth/0410079, [Erratum: Phys. Rev. Lett. 96, 039901 (2006)].
\bibitem{X.G. Huang}
X.G. Huang, P. Huovinen and X.N. Wang, Phys. Rev. C84 (2011) 054910, 1108.5649.
\bibitem{X.G. Huang1}
X.G. Huang, Rept. Prog. Phys. 79 (2016) 076302, 1509.04073.
\bibitem{F. Becattini}
F. Becattini and F. Piccinini, Ann. Phys. (Amsterdam) 323 (2008) 2452 .
\bibitem{F. Becattini1}
F. Becattini et al., Ann. Phys. (Amsterdam) 338 (2013) 32 .
\bibitem{F. Becattini2}
F. Becattini et al., Eur. Phys. J. C75 (2015) 406, 1501.04468.
\bibitem{A. Aristova}
A. Aristova et al., (2016), 1606.05882.
\bibitem{W.T. Deng}
W.T. Deng and X.G. Huang, Phys. Rev. C93 (2016) 064907, 1603.06117.
\bibitem{Shu Ebihara}
Shu Ebihara, Kenji Fukushima, Kazuya, Mameda, Phys. Lett. B
 764, 10 (2017).
\bibitem{L. Adamczyk et al}
L. Adamczyk et al. [STAR], Nature 548, 62-65 (2017) [arXiv:1701.06657 [nucl-ex]].
\bibitem{J. Adam et al}
J. Adam et al. [STAR], Phys. Rev. C 98, 014910 (2018) [arXiv:1805.04400 [nucl-ex]].
\bibitem{S. Acharya et al}
S. Acharya et al. [ALICE], Phys. Rev. C 101, 044611 (2020) [arXiv:1909.01281 [nucl-ex]].



\bibitem{Y. Tsue1}
Y. Tsue, J. da Provid\^{e}ncia, C. Provid\^{e}ncia, M. Yamamura,
and H. Bohr, Prog. Theor. Exp. Phys. 2013, 103D01 (2013).
\bibitem{Y. Tsue2}
Y. Tsue, J. da Provid\^{e}ncia, C. Provid\^{e}ncia, M. Yamamura,
and H. Bohr, Prog. Theor. Exp. Phys. 2015, 103D02 (2015).
\bibitem{Y. Tsue3}
Y. Tsue, J. da Provid\^{e}ncia, C. Provid\^{e}ncia, M. Yamamura,
and H. Bohr, Prog. Theor. Exp. Phys. 2015, 103D01 (2015).
\bibitem{H. Matsuoka1}
H. Matsuoka, Y. Tsue, J. da Provid\^{e}ncia, C. Provid\^{e}ncia, M.
Yamamura, and H. Bohr, Prog. Theor. Exp. Phys. 2016,
053D02 (2016).

\bibitem{H. Matsuoka2}
H. Matsuoka, Y. Tsue, J. da Provid\^{e}ncia, M. Yamamura,Phys. Rev. D 95, 054025 (2017).
\bibitem{X. G. Huang4}
X. G. Huang, J. Liao, Q. Wang, and X. L. Xia, Vorticity and
spin polarization in heavy ion collisions: Transport models,
arXiv:2010.08937.
\bibitem{F. Becattini6}
F. Becattini and M.A. Lisa, Polarization and Vorticity in the
quark gluon plasma, Annu. Rev. Nucl. Part. Sci. 70, 395
(2020).
\bibitem{Yu Guo}
Yu Guo, Jinfeng Liao, Enke Wang, Hongxi Xing, and Hui Zhang, Hyperon polarization from the vortical fluid in low-energy nuclear collisions, Phys. Rev. C 104, L041902.








\bibitem{S. Jeon and V. Koch}
S. Jeon and V. Koch, Event by event fluctuations, in Quark gluon plasma, R.C. Hwa and
X.N. Wang eds., World Scientific, pp. 430-490 [hep-ph/0304012] [INSPIRE].

\bibitem{V. Koch}
V. Koch, Hadronic fluctuations and correlations, arXiv:0810.2520 [INSPIRE].

\bibitem{A. Bzdak1}
A. Bzdak, V. Koch and J. Liao, Remarks on possible local parity violation in heavy ion
collisions, Phys. Rev. C 81 (2010) 031901 [arXiv:0912.5050] [INSPIRE].

\bibitem{STAR collaboration1}
STAR collaboration, X.-F. Luo, Probing the QCD critical point with higher moments of
net-proton multiplicity distributions, J. Phys. Conf. Ser. 316 (2011) 012003
[arXiv:1106.2926] [INSPIRE].

\bibitem{S. Gupta}
S. Gupta, X. Luo, B. Mohanty, H.G. Ritter and N. Xu, Scale for the phase diagram of
quantum chromodynamics, Science 332 (2011) 1525 [arXiv:1105.3934] [INSPIRE].

\bibitem{A. Bzdak2}
A. Bzdak, V. Koch and J. Liao, Charge-dependent correlations in relativistic heavy ion
collisions and the chiral magnetic effect, Lect. Notes Phys. 871 (2013) 503
[arXiv:1207.7327] [INSPIRE].


\bibitem{A. Bazavov et al.}
A. Bazavov et al., Freeze-out conditions in heavy ion collisions from QCD thermodynamics,
Phys. Rev. Lett. 109 (2012) 192302 [arXiv:1208.1220] [INSPIRE].

\bibitem{X.F. Luo1}
X.F. Luo, B. Mohanty, H.G. Ritter and N. Xu, Search for the QCD critical point: higher
moments of net-proton multiplicity distributions, Phys. Atom. Nucl. 75 (2012) 676
[arXiv:1105.5049] [INSPIRE].

\bibitem{X. Luo}
X. Luo, J. Xu, B. Mohanty and N. Xu, Techniques in the moment analysis of net-proton
multiplicity distributions in heavy-ion collisions, arXiv:1302.2332 [INSPIRE].

\bibitem{STAR collaboration2}
STAR collaboration, X. Luo, Search for the QCD critical point by higher moments of
net-proton multiplicity distributions at STAR, Nucl. Phys. A 904-905 (2013) 911c-914c
[arXiv:1210.5573] [INSPIRE].

\bibitem{Shuzhe Shi}
Shuzhe Shi and Jinfeng Liao, 10.1007/JHEP 06 (2013) 104.
















\bibitem{R. V. Gavai}
R. V. Gavai and S. Gupta, Phys. Rev. D 68 (2003) 034506.
\bibitem{R. V. Gavai1}
R. V. Gavai and S. Gupta, Phys. Rev. D 71 (2005) 114014.
\bibitem{S. Gupta10}
S. Gupta, N. Karthik and P. Majumdar, Phys. Rev. D 90 (2014) 034001.

\bibitem{L. Adamczyk1}
L. Adamczyk et al., (STAR Collaboration), Phys. Rev. Lett. 112, 032302 (2014).
\bibitem{L. Adamczyk2}
L. Adamczyk et al., (STAR Collaboration), Phys. Rev. Lett. 113, 092301 (2014).
\bibitem{X. Luo2}
X. Luo,  EPJ Web of Conferences, 141.04001 (2017).





\bibitem{R. Gavai and S. Gupta}
R. Gavai and S. Gupta, "Fluctuations, strangeness and quasi-quarks in heavy-ion collisions from lattice QCD," Phys. Rev. D 73, 014004 (2006).

\bibitem{S. Borsanyi1}
S. Borsanyi, Z. Fodor, S. Katz, S. Krieg, C. Ratti and K. Szabo, "Freeze-out parameters: lattice meets experiment," Phys. Rev. Lett. 111, 062005 (2013).

\bibitem{S. Borsanyi2}
S. Borsanyi, "Thermodynamics of the QCD transition from lattice," Nucl. Phys. A 904-905, 270c-277c
(2013).

\bibitem{R. Bellwied}
R. Bellwied, S. Borsanyi, Z. Fodor, S. Katz, A. Pasztor, C. Ratti and K. Szabo, "Fluctuations and
correlations in high temperature QCD," Phys. Rev. D 92, no.11, 114505 (2015).

\bibitem{H. T. Ding}
H.T. Ding, S. Mukherjee, H. Ohno, P. Petreczky and H.P. Schadler, "Diagonal and off-diagonal quark
number susceptibilities at high temperatures," Phys. Rev. D 92, no.7, 074043 (2015).







\bibitem{T. Kunihiro}
 T. Kunihiro, Phys. Lett. B 271 (1991) 395.
\bibitem{H. Fujii and M. Ohtani}
 H. Fujii and M. Ohtani, Phys. Rev. D 70 (2004) 014016.
\bibitem{Y. Hatta and T. Ikeda}
 Y. Hatta and T. Ikeda, Phys. Rev. D 67 (2003) 014028.
\bibitem{B. J. Schaefer and J. Wambach}
 B.J. Schaefer and J. Wambach, Phys. Rev. D 75 (2007) 085015.



\bibitem{A. R. Bodmer}
 A.R. Bodmer, Phys. Rev. D 4 (1971) 1601.
\bibitem{E. Witten}
 E. Witten, Phys. Rev. D 30 (1984) 272.
\bibitem{C. Alcock}
C. Alcock, E. Farhi, and A.V. Olinto, Astrophys. J. 310 (1986) 261.
\bibitem{C. Alcock1}
C. Alcock and A.V. Olinto, Ann. Rev. Nucl. Part. Sci. 38 (1988) 161.
\bibitem{J. Madsen}
J. Madsen, Lecture Notes in Physics 516 (1999) 162.
\bibitem{N. K. Glendenning1}
N.K. Glendenning and F. Weber, Astrophys. J. 400 (1992) 647.
\bibitem{N. K. Glendenning2}
N.K. Glendenning, Ch. Kettner, and F. Weber, Astrophys. J. 450 (1995) 253.
\bibitem{N. K. Glendenning3}
N.K. Glendenning, Ch. Kettner, and F. Weber, Phys. Rev. Lett. 74 (1995) 3519







\bibitem{H.Terazawa}
H. Terazawa, INS-Report-338 (INS, Univ. of Tokyo, 1979).
\bibitem{H.Terazawa1}
H. Terazawa, J. Phys. Soc. Jpn. 58 (1989) 3555.
\bibitem{H.Terazawa2}
H. Terazawa, J. Phys. Soc. Jpn. 58 (1989) 4388.
\bibitem{H.Terazawa3}
H. Terazawa, J. Phys. Soc. Jpn. 59 (1990) 1199.




\bibitem{Y. Jiang1}
Y. Jiang and J. Liao, Phys. Rev. Lett. 117, 192302 (2016).



\bibitem{M. Buballa}
M. Buballa, Phys. Rept. 407, 205 (2005).



\bibitem{J. I. Kapusta}
J. I. Kapusta, Finite Temperature Field Theory (Cambridge
University Press, Cambridge, England, 1989).


\bibitem{F. Becattinia}
F. Becattinia, F. Piccinini Annals of Physics 323 (2008) 2452-2473.


\bibitem{R.V. Gavai and S. Gupta1}
R.V. Gavai and S. Gupta, Pressure and nonlinear susceptibilities in QCD at finite chemical
potentials, Phys. Rev. D 68 (2003) 034506 [hep-lat/0303013] [INSPIRE].

\bibitem{R.V. Gavai and S. Gupta2}
R.V. Gavai and S. Gupta, Simple patterns for non-linear susceptibilities near Tc,
Phys. Rev. D 72 (2005) 054006 [hep-lat/0507023] [INSPIRE].

\bibitem{R.V. Gavai and S. Gupta3}
R.V. Gavai and S. Gupta, The critical end point of QCD, Phys. Rev. D 71 (2005) 114014
[hep-lat/0412035] [INSPIRE].

\bibitem{C.R. Allton et al.}
C.R. Allton et al., Thermodynamics of two flavor QCD to sixth order in quark chemical
potential, Phys. Rev. D 71 (2005) 054508 [hep-lat/0501030] [INSPIRE].

\bibitem{M. Cheng et al.}
M. Cheng et al., Baryon number, strangeness and electric charge fluctuations in QCD at high
temperature, Phys. Rev. D 79 (2009) 074505 [arXiv:0811.1006] [INSPIRE].

\bibitem{S. Bors anyi et al.}
S. Bors$\acute{a}$nyi et al., Fluctuations of conserved charges at finite temperature from lattice QCD,
JHEP 01 (2012) 138 [arXiv:1112.4416] [INSPIRE].

\bibitem{HotQCD collaboration}
HotQCD collaboration, A. Bazavov et al., Fluctuations and correlations of net baryon
number, electric charge and strangeness: a comparison of lattice QCD results with the hadron
resonance gas model, Phys. Rev. D 86 (2012) 034509 [arXiv:1203.0784] [INSPIRE].



\bibitem{H. L. Chen}
H.L. Chen, K. Fukushima, X. G. Huang, and K. Mameda, Phys. Rev. D 93, 104052 (2016).
\bibitem{S. Ebihara}
S. Ebihara, K. Fukushima, and K. Mameda, Phys. Lett. B
764, 94 (2017).
\bibitem{M. N. Chernodub1}
M.N. Chernodub and S. Gongyo, J. High Energy Phys. 01
(2017) 136 [arXiv:1611.02598] [INSPIRE].
\bibitem{M. N. Chernodub2}
M.N. Chernodub and S. Gongyo, Phys. Rev. D 95, 096006
(2017).
\bibitem{M. N. Chernodub3}
M.N. Chernodub, Inhomogeneous confining-deconfining phases in rotating plasmas, Phys. Rev. D 103, 054027 (2021) [arXiv:2012.04924] [INSPIRE].


\bibitem{Xinyang Wang}
Xinyang Wang,  Minghua Wei,  Zhibin Li and Mei Huang,  Quark matter under rotation in the NJL model with vector interaction, Phys. Rev. D 99, 016018 (2019).


\bibitem{H.Kohyama}
H. Kohyama, D.Kimura, T.Inagaki, Nuclear Physics B 906(2016) 524-548.

\bibitem{P.de Forcrand}
P. de Forcrand, PoS LAT2009,010 (2009), arXiv:1005.0539.





\bibitem{S. P. Klevansky}
S.P. Klevansky, Rev. Mod. Phys. 64, 649 (1992).








\bibitem{M. Buballa}
M. Buballa,  NJL-model analysis of dense quark matter. Phys. Rep. 407, 205-376 (2005).
\bibitem{H. Kohyama}
H. Kohyama,  D. Kimura,  T. Inagaki, Regularization dependence on
phase diagram in Nambu-Jona-Lasinio model. Nucl. Phys. B 896, 682-
715 (2015).
\bibitem{C. D.  Roberts}
C. D.  Roberts,  A. G. Williams, Dyson-Schwinger equations and their
application to hadronic physics. Prog. Part. Nucl. Phys. 33, 477-575
(1994).
\bibitem{R.  Alkofer}
R.  Alkofer,  L. Von Smekal, The infrared behaviour of QCD Green's
functions: Confinement, dynamical symmetry breaking, and hadrons as
relativistic bound states. Phys. Rep. 353, 281-465 (2001).





\bibitem{C. S. Fischer}
C. S. Fischer, Infrared properties of QCD from Dyson-Schwinger
equations. J. Phys. G32, R253-R291 (2006).
\bibitem{I. C.}
I. C.  Clo$\ddot{e}$t,  C. D. Roberts, Explanation and prediction of observables
using continuum strong QCD. Prog. Part. Nucl. Phys. 77, 1-69 (2014).
\bibitem{T.M. Schwarz}
T.M. Schwarz, S.P. Klevansky, G. Papp, The Phase diagram
and bulk thermodynamical quantities in the NJL model at finite
temperature and density. Phys. Rev. C 60, 055205 (1999).
arXiv:nucl-th/9903048 [nucl-th]
\bibitem{P. Zhuang}
P. Zhuang, M. Huang, Z. Yang, Density effect on hadronization
of a quark plasma. Phys. Rev. C 62, 054901 (2000).
arXiv:nucl-th/0008043 [nucl-th]
\bibitem{J.-W. Chen}
J. W. Chen, J. Deng, L. Labun, Baryon susceptibilities, non-Gaussian moments, and theQCDcritical point. Phys. Rev.D92(5),
054019 (2015). arXiv:1410.5454 [hep-ph]



\bibitem{J.-W. Chen1}
J. W. Chen, J. Deng, H.Kohyama, L. Labun, Robust characteristics
of nongaussian fluctuations from the NJL model. Phys. Rev. D
93(3), 034037 (2016). arXiv:1509.04968 [hep-ph]
\bibitem{W. Fan}
W. Fan, X. Luo, H.-S. Zong, Mapping the QCD phase diagram
with susceptibilities of conserved charges within Nambu-Jona-
Lasinio model. Int. J. Mod. Phys. A 32(11), 1750061 (2017).
arXiv:1608.07903 [hep-ph]
\bibitem{W. Fan1}
W. Fan, X. Luo, H. Zong, Identifying the presence of the critical
end point in QCD phase diagram by higher order susceptibilities.
arXiv:1702.08674 [hep-ph]
\bibitem{W.-J. Fu}
W. J. Fu, Y. L. Wu, Fluctuations and correlations of conserved
charges near the QCD critical point. Phys. Rev. D 82, 074013
(2010). arXiv:1008.3684 [hep-ph]



\bibitem{E.S. Bowman}
E.S. Bowman, J.I.Kapusta,Critical points in the linear sigmamodel
with quarks. Phys. Rev. C 79, 015202 (2009). arXiv: 0810.0042
[nucl-th]
\bibitem{H. Mao}
H. Mao, J. Jin, M. Huang, Phase diagram and thermodynamics of
the Polyakov linear sigma model with three quark flavors. J. Phys.
G37, 035001 (2010). arXiv: 0906.1324 [hep-ph]
\bibitem{B.J. Schaefer}
B. J. Schaefer, M. Wagner, QCD critical region and higher
moments for three flavor models. Phys. Rev. D 85, 034027 (2012).
arXiv:1111.6871 [hep-ph]
\bibitem{B.-J. Schaefer1}
B. J. Schaefer, M. Wagner, Higher-order ratios of baryon number
cumulants. Cent. Eur. J. Phys. 10, 1326-1329 (2012).
arXiv:1203.1883 [hep-ph]
\bibitem{S.-X. Qin}
S. X. Qin, L. Chang, H. Chen, Y.-X. Liu, C.D. Roberts, Phase
diagram and critical endpoint for strongly-interacting quarks. Phys.
Rev. Lett. 106, 172301 (2011). arXiv: 1011.2876 [nucl-th]


\bibitem{J. Luecker}
J. Luecker, C.S. Fischer, L. Fister, J.M. Pawlowski, Critical
point and deconfinement from Dyson-Schwinger equations.
PoSCPOD2013, 057 (2013). arXiv:1308.4509 [hep-ph]
\bibitem{W.-J. Fu1}
W. J. Fu, J.M. Pawlowski, F. Rennecke, B.-J. Schaefer, Baryon
number fluctuations at finite temperature and density. Phys. Rev.
D 94(11), 116020 (2016). arXiv: 1608.04302 [hep-ph]
\bibitem{Luis A. H. Mamani}
Luis A. H. Mamani, Cesar V. Flores, and Vilson T. Zanchin. Phys. Rev. D 102, 066006 (2020).


\bibitem{J.N. Guenther}
J.N. Guenther, Overview of the QCD phase diagram-Recent progress from the lattice, Eur.Phys. J. A 57, 136 (2021).




\bibitem{Victor E. Ambrus}
Victor E. Ambrus, Elizabeth Winstanley, Phys. Rev. D 93, 104014 (2016).
\bibitem{C. Ratti}
C. Ratti, M.A. Thaler and W. Weise, Phys. Rev. D 73, 014019 (2006).



\end{thebibliography}
\end{document}